\documentclass[aps,prb,twocolumn]{revtex4}  %,,preprint
\usepackage{graphicx}% Include figure files
\usepackage{dcolumn}% Align table columns on decimal point
\usepackage{bm}% bold math  \bm{\mu}
\usepackage{amsmath}
\newcommand{\comment}[1]{}

\begin{document}
\renewcommand{\theequation}{\arabic{section}.\arabic{equation}}
%\draft
%\narrowtext
%\twocolumn
%\pagestyle{myheadings} \markboth{}{}

%\tableofcontents
%%%%%%%%%%%%%%%%%%%%%%%%%%%%%%%%%%%%%%%%%%%%%%%%%%%%%%%%%%%%%%%%%%%%%%%%
%%%%%%%%%%%%%%%%%%%%%%%%%%%%%%%%%%%%%%%%%%%%%%%%%%%%%%%%%%%%%%%%%%%%%%%%%%%%%%

\title{Shear Thinning in Lennard-Jones Fluids
by  Stochastic Dissipative Molecular Dynamics Simulation}

\author{Phil Attard}
\email[]{phil.attard1@gmail.com}
%\thanks{}

%\affiliation{phil.attard1@gmail.com}

%\date{\today. Begun 9 February, 2018.  Papers/Current/Shear/Shear.tex }

\begin{abstract}
The shear viscosity of a Lennard-Jones fluid is obtained
by stochastic dissipative molecular dynamics (SDMD) simulations.
%which equations of motion preserve
%the non-equilibrium probability distribution.
A generic constraint to the equations of motion is given
that reduces the sensitivity of
the  shear viscosity to the value
of the the fluctuation-dissipation or thermostat parameter.
At high shear rates
the shear viscosity becomes dependent on the size of the system,
and corrections to the equipartition kinetic temperature arise.
At constant kinetic temperature the shear viscosity
is shown to decrease with increasing shear rate.
\end{abstract}

\maketitle
%%%%%%%%%%%%%%%%%%%%%%%%%%%%%%%%%%%%%%%%%%%%%%%%%%%%%%%%%%%%%%%%%%%%%%%%%%

%%%%%%%%%%%%%%%%%%%%%%%%%%%%%%%%%%%%%%%%%%%%%%%%%%%%%%%%%%%%%%%%%%%%%%%%%%
%
\section{Introduction}
\setcounter{equation}{0} \setcounter{subsubsection}{0}
%
%%%%%%%%%%%%%%%%%%%%%%%%%%%%%%%%%%%%%%%%%%%%%%%%%%%%%%%%%%%%%%%%%%%%%%%%%%

The value of computer simulations is that
they provide molecular-level detail,
and quantitative results that are completely reliable.
In principle the only approximations invoked
are the finite-size of the system
and the finite statistical sampling of the configurations,
both of which which can be increased
to improve or check the accuracy of the results.\cite{Allen87}

In practice the situation can be more complicated,
and it is debatable whether simulations are completely reliable
for non-equilibrium systems,
since for these cases
there is little agreement on the foundations or formulation
of statistical mechanics or thermodynamics.
Contention continues about fundamental matters
such as the form of the probability distribution,
the nature of the appropriate equations of motion,
and the physical origin and meaning of various thermostats.
Whilst in the linear regime
different non-equilibrium simulation methods give consistent results,
in the non-linear regime there is
considerable uncertainty and quantitative disagreement
between different computer simulation methods.

For example,
the first computer simulations of shear flow in simple atomic fluids
were carried out over 30 years ago,\cite{Erpenbeck84,Heyes86}
but the many papers on the subject since then
have produced contradictory results in the high shear rate regime,
with both shear thickening, and shear thinning observed
(see, for example, Refs.~[\onlinecite{Todd07,Hoover08}]
and references therein),
as well as sensitivity to the particular thermostat employed.
\cite{Hoover08,Khare97,Berro11,Delhommelle03}
Even the structure of shearing simple fluids has been disputed,
with some simulations supporting the existence of a string phase
at high shear rates,
\cite{Erpenbeck84,Loose89,Loose92,Bagchi96}
and others showing the opposite.
\cite{Evans86,Padilla95,Travis96,Delhommelle03}
%Although there are probably more results for shear thinning then thickening,
%Even more confusing, an investigation aiming to sort out the issue
%showed an initial shear thinning, followed by strong shear thickening.
%\cite{Delhommelle03}

The non-equilibrium molecular dynamics (NEMD) method dominates the field,
but in contrast to equilibrium simulations
it is arguable (see Ref.~[\onlinecite{NETDSM}] and below)
that this is based on five explicit approximations
of contentious validity:
(1) The equations of motion
are created artificially to give the desired linear flow,
but they have no thermodynamic justification or
demonstrated validity in the non-linear regime.
(2) The thermostats that have been used likewise have no thermodynamic basis
(arguments based on a local Lagrangian co-moving reference frame
neglect the effects of any velocity gradient),
and they variously invoke a non-physical extra phase space variable,
or an artificial constraint on the kinetic energy.
(3) Every thermostat relies upon a formula for the local temperature,
most usually the equipartition theorem or
Rugh's variant,
which formulae are invalid in the non-linear non-equilibrium regime
(see Eqs~(\ref{Eq:EquiThm}) and (\ref{Eq:EquiThm-Shear}) below).
(4) Applying a thermostat assumes that the temperature should be uniform,
which need not in fact be the case in a non-equilibrium system
(see Eq.~(\ref{Eq:T(z)}) below).
(5) The stochastic contribution to the equations of motion,
which necessarily results from the projection of the reservoir interactions,
is neglected.\cite{NETDSM}
The effects of these five approximations
are tolerable or negligible in the linear regime,
but their unknown consequences in the non-linear regime
undermine confidence in the results.
Indeed, Hoover,
one of the inventors of NEMD,
finds that  at high shear rates
the DOLLS and SLLOD equations of motion for shear flow
quantitatively and even qualitatively disagree
with each other
and with a more physical model of boundary driven flow.\cite{Hoover08}

This paper presents non-equilibrium computer simulation results
for a shearing Lennard-Jones fluid.
The concrete aim is to obtain the shear viscosity
as a function of shear rate,
and to establish as reliably as possible
whether or not a simple fluid is shear thinning or shear thickening.
More broadly,
the analysis and results shed light on some of the reasons
for the discrepancies in previous simulations at high shear rates.
It is found analytically and numerically
that a shearing fluid is inhomogeneous in density and temperature,
and consequently that the shear viscosity can be dependent
on the system size (as also is the pressure).
It is also found that  the viscosity can be sensitive
to the value of the fluctuation-dissipation thermostat parameter,
which sensitivity can be reduced by adding a constraint
to the equations of motion.
Finally,
it is found that whether the fluid is shear thinning or shear thickening
depends to some extent on whether the shear flow is boundary driven
or else uniformly driven,
and on the sign of the temperature derivative of the viscosity.

Of course one can legitimately ask whether the present results
are any more reliable in the non-linear regime
than previous NEMD and other results.
\cite{Erpenbeck84,Heyes86,Todd07,Hoover08,Khare97,Berro11,Delhommelle03,%
Loose89,Loose92,Bagchi96,Evans86,Padilla95,Travis96}
This question is addressed in the conclusion,
but here it may simply be pointed out
that the present stochastic dissipative molecular dynamics (SDMD) algorithm
is a representation of the formally exact transition probability
of a non-equilibrium system,
and as such it preserves the non-equilibrium probability in phase space.
\cite{NETDSM,STD2,AttardV}
The SDMD method has previously been tested for
conductive heat flow\cite{AttardV}
and for driven Brownian motion.\cite{AttardIX}
For the present case of shear flow,
the SDMD approach is essentially the same as the stochastic Langevin equation,
which has previously been used for molecular dynamics simulations
of complex shear flow, particularly of bead polymers
(see Refs~[\onlinecite{Kremer90,Lyulin99,Shang17}]
and references therein).
In  contrast to the NEMD and similar methods,
the SDMD algorithm avoids artificial equations of motion,
and unjustified thermostats.
Neither does it have to postulate
an equilibrium functional form
nor a particular local value for the temperature.
Finally, and uniquely,
it explicitly accounts for the projected reservoir interactions.

In principle the SDMD results are exact,
but again the situation in practice is more complicated.
Even with the constraint introduced below,
large values of the fluctuation-dissipation parameter
still affect the results,
which creates uncertainty at high shear rates.
There is also a certain conceptual or physical uncertainty
that arises from the inhomogeneous nature
of the shearing system, and from the size-dependence of the viscosity,
which lead to questions regarding whether a boundary driven model
or a uniform shearing model is most appropriate.
These limit confidence in the results at very high shear rates,
and are discussed in more detail in the conclusion.
Nevertheless, the results at moderate shear rates
are sufficiently clear to conclude
that departing the linear regime Lennard-Jones fluids
are shear thinning.

% end comment %%%%%%%%%%%%%%%%%%%%%%%%%%%%%%%%%%%%%%%%%%

%%%%%%%%%%%%%%%%%%%%%%%%%%%%%%%%%%%%%%%%%%%%%%%%%%%%%%%%%%%%%%%%%%%%%%%%%%
%
\section{Equations of Motion and Simulation Details}
\setcounter{equation}{0} \setcounter{subsubsection}{0}
%
%%%%%%%%%%%%%%%%%%%%%%%%%%%%%%%%%%%%%%%%%%%%%%%%%%%%%%%%%%%%%%%%%%%%%%%%%%

\subsection{General Formulation}

In general, for a non-equilibrium system,
the stochastic, dissipative equations of motion
in phase space are
\cite{NETDSM,STD2,AttardIX}
\begin{eqnarray}
{\bf q}(t+\Delta_t) & = &
{\bf q}(t) + \frac{\Delta_t}{m} {\bf p}(t)
\nonumber \\
{\bf p}(t+\Delta_t) & = &
{\bf p}(t) + \Delta_t \dot{\bf p}^0(t)
+ {\bf R}(t).
\end{eqnarray}
Here $ {\bf q} = \{ {\bf q}_1, {\bf q}_2, \ldots , {\bf q}_N\}$
are the positions,
with
$ {\bf q}_j = \{ {q}_{jx}, {q}_{jy},  { q}_{jz}\}$,
and $ {\bf p} = m \dot {\bf q}$
are the momenta,
with $m$ being each particle's mass.
The adiabatic derivative of the momentum
is given by Hamilton's equations,
$\dot{\bf p}^0 = -\partial {\cal H}/\partial {\bf q}$.
These equations preserve the non-equilibrium
phase space probability density to linear order in the time step $\Delta_t$,
and in the fluctuation-dissipation parameter $\theta$.

The reservoir force (times the time step)
comprises dissipative and stochastic parts,
${\bf R}(t) = \overline {\bf R}(t) + \tilde {\bf R}(t)$.
The dissipative part is\cite{STD2}
\begin{equation} \label{Eq:olR}
\overline {\bf R}(t)
=
\frac{ \theta |\Delta_t| }{2k_\mathrm{B}}
\nabla_p S_\mathrm{r,st}({\bf \Gamma})
+ \frac{ \theta |\Delta_t | }{2k_\mathrm{B}}
[ \widehat t - 1 ] \overline {\nabla_p S_\mathrm{r,st}} ,
\end{equation}
and the stochastic part is Gaussian distributed with zero mean and variance
\begin{equation} \label{Eq:tildeR}
\left< \tilde {\bf R}(t')\tilde {\bf R}(t) \right>
=
\theta | \Delta_t | \delta(t-t') \underline{\underline{\mathrm I} }.
\end{equation}
In these $\widehat t = \mbox{sign} (\Delta_t)$,
and $\theta$ is the fluctuation-dissipation parameter.
(A constraint may be added to the dissipative force
to ameliorate the direct influence of the reservoir
on the transport coefficient;
see \S \ref{Sec:GenCon} below.)

The total entropy is the sum of the sub-system entropy
and the reservoir entropy.
For a point in phase space,
the sub-system entropy is zero,
$S_\mathrm{s}({\bf \Gamma}) = 0$.\cite{TDSM}
For a non-equilibrium system,
the reservoir (hence total) entropy
comprises static and dynamic parts,
$S_\mathrm{r}({\bf \Gamma},t)
=  S_\mathrm{r,st}({\bf \Gamma},t)+ S_\mathrm{r,dyn}({\bf \Gamma},t)$.
Obviously, the phase space probability
for the non-equilibrium system is
$\wp({\bf \Gamma},t) =
Z(t)^{-1} e^{ S_\mathrm{r}({\bf \Gamma},t) /k_\mathrm{B}}$.
\cite{NETDSM,STD2,AttardV}

The static part of the reservoir entropy
is the entropy that would be written down
if the sub-system were instantaneously an equilibrium system.
If the exchangeable linear additive variables are
${\bf X}$, and the conjugate field variables are ${\bf x}$,
then for the reservoir
${\bf x}_\mathrm{r} = \partial S_r/\partial {\bf X}_\mathrm{r}$.
In the usual Gibbsian fashion,
the static part of the reservoir entropy is
\begin{equation}
S_\mathrm{r,st}({\bf \Gamma},t)
= -{\bf X}_\mathrm{s}({\bf \Gamma},t) \cdot {\bf x}_\mathrm{r} .
\end{equation}
This procedure is concretely illustrated with shear flow below.

The dynamic part of the the reservoir  entropy is
\cite{NETDSM,STD2,AttardV}
\begin{equation}
S_\mathrm{r,dyn}({\bf \Gamma},t)
=
-\int_{-\infty}^t \mathrm{d} t'\;
\dot S_\mathrm{r,st}^0(\overline{\bf \Gamma}(t'|{\bf \Gamma},t),t') ,
\end{equation}
where in the integrand appear
the adiabatic time derivative of the static reservoir entropy,
and also the most likely trajectory from the current point.
The dynamic part of the reservoir entropy is not required explicitly
in the present paper.

The fluctuation-dissipation parameter $\theta$
in Eqs~(\ref{Eq:olR}) and (\ref{Eq:tildeR})
can be regarded as a thermodynamic drag coefficient
that occurs in the transition probability.
Its value cannot be too large
since the stochastic dissipative equations of motion
are based upon an expansion to linear order in $\theta$,
but it is otherwise arbitrary.
The essential point is that it must link
the dissipative and stochastic forces exactly as
in Eqs~(\ref{Eq:olR}) and (\ref{Eq:tildeR}).
This is the fluctuation-dissipation theorem,
which says in essence that the proportionality constant
is the same for both types of forces;
dissipation without fluctuation is forbidden.\cite{NETDSM,STD2}
(The violation of this principle
is a significant weakness of the NEMD method
and of other molecular dynamics algorithms\cite{Todd07,Hoover08}
that use solely a deterministic thermostat.)
The fluctuation-dissipation theorem guarantees the preservation
of the non-equilibrium probability in phase space,
and hence that time averages equal phase space averages.

The fluctuation-dissipation parameter is most simply taken to be a constant,
$\theta$.
%(This may also be loosely called the thermostat parameter,
%although it is important to emphasize
%that the reservoir creates both dissipative and stochastic forces.)
However, the analysis remains formally exact if it varies with particle,
$\theta_{i}$.
As a large class of non-equilibrium systems consist of boundary driven flow,
including the present shear flow,
it is useful to take this to be dependent on position,
$\theta_{i} = \theta({\bf q}_i)$.
Since this parameter represents the strength of the interaction
between the reservoir and the sub-system,
it is sensible to make this parameter large near the boundary between the two,
and small in the interior of the sub-system.
In this work, where the reservoir boundaries are located at $z = \pm L_z/2$,
one functional form that is explored below is
\begin{equation}
\theta({\bf r}) =  \left[ \frac{2 z}{L_z} \right]^{2n} \theta
,\;\; n = 0, 1, 2, \ldots
\end{equation}
For $n=0$ this is constant and the influence of the fluctuation-dissipation
parameter is uniform throughout the sub-system.
As $n$ is increased, its direct influence is increasingly confined to the
boundary region.

A second form that is explored below is
\begin{equation}
\theta({\bf r}) =
\left\{
\begin{array} {ll}
0 , & | z| \le L_z/2 - R_\mathrm{cut}, \\
\theta, & | z| > L_z/2 - R_\mathrm{cut} .
\end{array} \right.
\end{equation}
This confines the direct effect of the reservoir
to slabs of width $R_\mathrm{cut}$ next to each boundary.
In this work $R_\mathrm{cut}$ was chosen to be the same
as the cut-off to the pair potential,
but this is not essential.
This form for the fluctuation-dissipation parameter
will be referred to as the slab form, or the slab thermostat.
Periodic boundary conditions are used for $|z| > L_z/2$.
(In the case of shear flow, these are Lees-Edwards sliding brick
boundary conditions.)\cite{Lees72}

The slab form of position dependence
is similar to the simulation geometry
used by Khare \emph{et al.}\cite{Khare97}
and by Hoover \emph{et al.},\cite{Hoover08}
namely the central region evolves adiabatically via Hamilton's equations
of motion.
However it does differ in the boundary regions,
where in the present case the above equations of motion are used,
which means that the particles in the slab undergo
reservoir-induced shear flow.
%In the case of Hoover \emph{et al.}\cite{Hoover08}
%the atoms in the slabs are thermostatted
%and tethered to a rigid cubic lattice in uniform motion.
%It will be demonstrated in the results below
%that the simulated shear viscosity of the central part
%of the sub-system is sensitive to the treatment
%of the boundary region.

%%%%%%%%%%%%%%%%%%%%%%%%%%%
\subsubsection{General Constraint} \label{Sec:GenCon}

In non-equilibrium thermodynamics,
a quantity that often occurs
is the adiabatic rate of change of the static part of the reservoir entropy,
$\dot S_\mathrm{r,st}^0({\bf \Gamma},t)$.
The autocorrelation function of this
is essentially the Green-Kubo relation for the hydrodynamic
transport coefficients.\cite{NETDSM}
This quantity represents the adiabatic relaxation
of relevant macroscopic fluctuations.
The adiabatic dynamics can differ significantly
when the system is perturbed by the reservoir.
In general the reservoir influence affects the dynamic order
that this represents,
with the result that the transport coefficients
can be expected to depend on its value.
This expectation has been confirmed for the thermal conductivity
\cite{AttardIX}
and will be demonstrated for the shear viscosity below.
%The same sensitivity occurs for the thermostats that other workers
%have applied, as is also discussed in the results below.

As is discussed in \S 8.3 of Ref.~[\onlinecite{NETDSM}],
a general way of circumventing the problem
is to impose a constraint on the reservoir force
so that it is orthogonal to the gradient of the adiabatic flux,
\begin{equation}
\overline {\bf R} \cdot \nabla_p \dot S_\mathrm{r,st}^0({\bf \Gamma},t)
= 0.
\end{equation}
If $\overline {R}_{i\alpha}^\mathrm{f}$
is the $i\alpha$ component of the unconstrained dissipative force given above,
then one can introduce a Lagrange multiplier $\mu$
so that the constrained dissipative force is
\begin{equation}
\overline {R}_{i\alpha}
=
\overline {R}_{i\alpha}^\mathrm{f}
+ \mu \theta_{i}
\partial_{p i\alpha}  \dot S_\mathrm{r,st}^0   .
\end{equation}
The Lagrange multiplier is given by
\begin{equation}
\mu({\bf \Gamma},t)
=
\frac{-\sum_{i,\alpha}
\overline {R}_{i\alpha}^\mathrm{f}
\partial_{p i\alpha} \dot S_\mathrm{r,st}^0
}{
\sum_{i,\alpha} \theta_{i}
(\partial_{p i\alpha}  \dot S_\mathrm{r,st}^0)^2
}.
\end{equation}

Three points can be made about this procedure.
First, there is no need to add a similar constraint
to the stochastic part of the reservoir force because it averages to zero,
as has been confirmed numerically.
Second,
this constraint effectively projects the dissipative force onto a hypersurface
of dimension $3N-1$.
Since this is negligibly different to the dimensions of the unconstrained
dissipative force, imposing it should have negligible effect
on the preservation of the non-equilibrium phase space probability density
by the equations of motion.
And third,
if any particle is in a region such that
$\theta_i
%\equiv\theta({\bf q}_{i})
=0$,
then the unconstrained $\overline {R}_{i\alpha}^\mathrm{f}$,
constrained $\overline {R}_{i\alpha}$,
and stochastic $\tilde {R}_{i\alpha}$ reservoir forces are also zero,
which means that the particle moves adiabatically.

%%%%%%%%%%%%%%%%%%%%%%%%%%%
\subsubsection{Equipartition Theorem}

The equipartition theorem,
which is formally exact for a non-equilibrium system,
is most generally written as\cite{NETDSM,STD2}
\begin{equation}
\left< \nabla \nabla S_\mathrm{r}({\bf \Gamma},t) \right>
=
- k_\mathrm{B}^{-1}
\left< [\nabla S_\mathrm{r}({\bf \Gamma},t)] \,
[\nabla S_\mathrm{r}({\bf \Gamma},t)] \right> .
\end{equation}
This holds for the position components if, and only if,
there is a confining potential that makes the density vanish
at and beyond the boundary of the sub-system.
If one restricts this to momentum components,
takes the trace,
and neglects the fluctuations in the consequent extensive quantity,
then on the likely points of phase space this becomes
\begin{equation}
\nabla_p^2 S_\mathrm{r}({\bf \Gamma},t)
=
- k_\mathrm{B}^{-1}
[\nabla_p S_\mathrm{r}({\bf \Gamma},t)] \cdot
[\nabla_p S_\mathrm{r}({\bf \Gamma},t)] .
\end{equation}
For the canonical equilibrium system,
$ S_\mathrm{r}({\bf \Gamma},t) = -{\cal H}({\bf \Gamma})/T$.
Inserting this yields the equilibrium equipartition theorem,
$ 3N m k_\mathrm{B}T = {\bf p}\cdot {\bf p}$.

As mentioned above,
the reservoir entropy of a non-equilibrium system
is composed of static and dynamic parts,
$S_\mathrm{r}({\bf \Gamma},t) =
S_\mathrm{r,st}({\bf \Gamma},t) + S_\mathrm{r,dyn}({\bf \Gamma},t)$.
By definition, since the static part corresponds to
an instantaneous equilibrium system
it must have even parity with regard to velocity reversal.
Since this is arguably the dominant contribution,
one need only retain the odd projection
of the dynamic part of the reservoir entropy,
$S_\mathrm{r,dyn}({\bf \Gamma},t) \Rightarrow
S_\mathrm{r,dyn}^\mathrm{odd}({\bf \Gamma},t)$.\cite{NETDSM,STD2}
Since the dynamic part vanishes in the equilibrium case,
say $\gamma = 0$,
it must be at least linear in the non-equilibrium parameter,
$ S_\mathrm{r,dyn}^\mathrm{odd} \sim {\cal O}(\gamma)$,
$\gamma \rightarrow 0$.
(The non-equilibrium parameter $\gamma$ can characterize
the strength of an applied thermodynamic gradient,
or of an external time-varying potential,
as generic examples.)
If one restricts attention to dyadic elements of the same parity,
say $pp$ to be definite,
then the left hand side of the
generalized equipartition theorem in the non-equilibrium case
is
\begin{equation}
\left< \nabla_p \nabla_p S_\mathrm{r}({\bf \Gamma},t) \right>_\mathrm{ne}
=
\left< \nabla_p \nabla_p S_\mathrm{r,st}({\bf \Gamma},t) \right>_\mathrm{st}
+ {\cal O}(\gamma^2),
\end{equation}
and the right hand side is
\begin{eqnarray}
\lefteqn{
- k_\mathrm{B}^{-1}
\left< [\nabla_p S_\mathrm{r}({\bf \Gamma},t)] \,
[\nabla_p S_\mathrm{r}({\bf \Gamma},t)] \right>_\mathrm{ne}
}  \\
& = &
- k_\mathrm{B}^{-1}
\left< [\nabla_p S_\mathrm{r,st}({\bf \Gamma},t)] \,
[\nabla_p S_\mathrm{r,st}({\bf \Gamma},t)] \right>_\mathrm{st}
\nonumber \\ && \mbox{ }
- k_\mathrm{B}^{-1}
\left< [\nabla_p S_\mathrm{r,dyn}^\mathrm{odd}({\bf \Gamma},t)] \,
[\nabla_p S_\mathrm{r,dyn}^\mathrm{odd}({\bf \Gamma},t)] \right>_\mathrm{st}
+ {\cal O}(\gamma^2)
\nonumber \\ & = &
- k_\mathrm{B}^{-1}
\left< [\nabla_p S_\mathrm{r,st}({\bf \Gamma},t)] \,
[\nabla_p S_\mathrm{r,st}({\bf \Gamma},t)] \right>_\mathrm{st}
+ {\cal O}(\gamma^2).\nonumber
\end{eqnarray}
A similar result holds for $qq$ elements.
Taking the trace of this
(which makes it extensive),
and neglecting fluctuations
(because for an extensive quantity they are relatively negligible),
gives on the likely points in phase space
\begin{eqnarray} \label{Eq:EquiThm}
\lefteqn{
\nabla_p^2 S_\mathrm{r,st}({\bf \Gamma},t)
}  \\
& = &
- k_\mathrm{B}^{-1}
[\nabla_p S_\mathrm{r,st}({\bf \Gamma},t)] \cdot
[\nabla_p S_\mathrm{r,st}({\bf \Gamma},t)]
+ {\cal O}(\gamma^2).\nonumber
\end{eqnarray}
This is the non-equilibrium analogue
of the usual equilibrium equipartition theorem;
in both cases the left hand side is proportional to the reservoir temperature.
%Applied to shear flow,
%this result relates the peculiar kinetic energy to the temperature
%(see below).

It is important to emphasize that this result is only correct
to zeroth and linear order in the non-equilibrium parameter.
The point is that many NEMD simulation algorithms\cite{Todd07,Hoover08}
invoke the equipartition theorem to obtain
the temperature of the configuration from the peculiar kinetic energy,
which is then used in the thermostat.
Such an argument is based on a locally co-moving Lagrangian reference frame.
However, since it neglects the contribution from the velocity gradient,
this expression cannot be used
for strongly non-equilibrium systems
because it is not exact beyond leading order.
%The effects of applying this form of the equipartition theorem
%beyond its regime of validity are unknown.
This is  one of the reasons
for the uncertainty in existing simulations
for the behavior of the shear viscosity at high shear rates.

Beyond the equipartition theorem,
Rugh
has given a rather general expression for the temperature
as a configurational average for an isolated, equilibrium system.
\cite{Rugh97,TDSM}
This has also been used in NEMD simulations,\cite{Delhommelle03}
even though there is no justification for it as a non-equilibrium average.
In fact, since the usual equilibrium equipartition theorem
is a special case of Rugh's expression,
the result given above
demonstrates that  Rugh's expression cannot be valid beyond
leading order in the non-equilibrium parameter.
(A critical discussion of the application of a Rugh temperature
to non-equilibrium systems is also given by
 Hoover and Hoover.)\cite{Hoover09}

%One further difficulty with applying Rugh's expression,
%or indeed any expression for the sub-system temperature,
%to a non-equilibrium system,
%is that the thermostat seeks to fix the local temperature
%equal to that specified by the reservoir.
%In fact however, for boundary driven thermodynamic flows,
%such as shear flow that is discussed shortly,
%the local temperature of the sub-system is \emph{not} equal
%to that of the reservoir.
%So even if one did have a reliable way to measure
%the temperature of a configuration of the sub-system,
%it may not be appropriate to apply a thermostat
%that applies the reservoir temperature everywhere.

Beyond the above criticism
(that neither the equipartition theorem nor Rugh's expression
are applicable beyond linear order in the non-equilibrium parameter),
it should also be pointed out that the left hand side gives
the reservoir temperature,
not the sub-system temperature.
In an equilibrium system these are equal, of course,
and one can regard the equipartition theorem as giving
the sub-system temperature at each phase space point.
Because it is an equilibrium system,
one can further demand that this sub-system configurational temperature
be uniform and equal to the reservoir temperature,
which makes it useful as a thermostat control function.
In a non-equilibrium system neither property is true.
One cannot assume that the sub-system temperature is equal to the
reservoir temperature,
and one cannot assume that the sub-system temperature
is uniform throughout the sub-system.
Both assumptions become increasingly dubious
as the non-equilibrium parameter is increased,
and they create doubts about any thermostatted  equations of motion
in the non-linear regime.
(The present SDMD equations of motion
invoke only the reservoir temperature,
and have no need for a sub-system configurational temperature,
and make no assumptions regarding its homogeneity.)

%%%%%%%%%%%%%%%%%%%%%%%%%%%
\subsection{Shear Flow}

%%%%%%%%%%%%%%%%%%%%%%%%%%%
\subsubsection{Isolated System}

The velocity is the thermodynamic conjugate of the momentum
(see \S 9.6 of Ref.~\onlinecite{NETDSM}),
\begin{equation}
\frac{\partial \sigma({\bf r})} {\partial {\bf p}({\bf r})}
= \frac{-1}{T({\bf r})} {\bf v}({\bf r}) .
\end{equation}
Here $\sigma({\bf r})$ is the (first) entropy density
of an isolated system,
${\bf p}({\bf r}) = m n({\bf r}) {\bf v}({\bf r})$,
is the momentum density,
$m$ is the particle mass,
$n({\bf r}) $ is the number density,
${\bf v}({\bf r}) $ is the velocity,
and $T({\bf r})$ is the temperature.

Based on this,
for shear flow one formulates macrostates as specified values
of the zeroth and first sub-system momentum moments,
to which the zeroth and first sub-system velocities
(see \S \ref{Sec:Res} below for their interpretation)
are thermodynamically conjugate,\cite{NETDSM}
\begin{equation}
{\bf v}_{\mathrm{s}0}
\equiv
-T
\frac{\partial S_\mathrm{s}({\bf P}_{0},\underline{\underline {\mathrm P}}_1)
}{\partial {\bf P}_{0}}
\mbox{ and }
\underline{\underline {\mathrm v}}_{\mathrm{s}1}
\equiv
-T
\frac{\partial S_\mathrm{s}({\bf P}_{0},\underline{\underline {\mathrm P}}_1)
}{\partial \underline{\underline {\mathrm P}}_1}.
\end{equation}
The other state variables are not shown explicitly.
The zeroth and first momentum moments are
\begin{equation}
{\bf P}_{0}   \equiv  \int_V  \mathrm{d}{\bf r} \, {\bf p}({\bf r})
\mbox{ and }
\underline{\underline {\mathrm P}}_1  \equiv
\int_V \mathrm{d}{\bf r} \, {\bf r} {\bf p}({\bf r}) .
\end{equation}

From momentum conservation,
the rate of change of the momentum density is just
the negative divergence of the flux.
Hence
the adiabatic rate of change of the first momentum moment tensor is
\begin{eqnarray}
\dot{\underline{ \underline {\mathrm P}}}_1^0
& = &
\int_V \mathrm{d}{\bf r} \, {\bf r} \dot {\bf p}^0({\bf r})
\nonumber \\ & = &
- \int_V \mathrm{d}{\bf r} \, {\bf r}
\nabla \cdot \underline{ \underline {\mathrm J}}({\bf r})
\nonumber \\ & = &
-\oint_A \mathrm{d}{\bf r} \,
{\bf r} \underline{ \underline {\mathrm J}}({\bf r})
+
\int_V \mathrm{d}{\bf r} \,\underline{ \underline {\mathrm J}}({\bf r})
\nonumber \\ & = &
V \underline{ \underline {\mathrm J}} .
\end{eqnarray}
This assumes that the momentum flux is uniform over the volume
and vanishes on the boundary.

In general
the total momentum flux tensor consists of the pressure tensor
and the convective  momentum flux,
$\underline{ \underline {\mathrm J}} =
\underline{\underline {\mathrm P}} + m n {\bf v} \, {\bf v}$
(see \S 5.1.4 of Ref.~\onlinecite{NETDSM}).
The pressure tensor comprises the thermodynamic pressure
and the viscous pressure tensor,
$\underline{\underline {\mathrm P}} =
p \underline{ \underline {\mathrm I}} + \underline{ \underline \Pi}$,
with the latter also known as the diffusive momentum flux,
$\underline{ \underline \Pi} =\underline{ \underline {\mathrm J}}^0 $.
%The component $J_{xz} = J_{zx} $ is the $x$-momentum
%per unit area per unit time
%crossing a plane perpendicular to the $z$-axis.
One sees therefore that the adiabatic rate of change of the
first momentum moment with the convective  momentum flux removed
is proportional to the pressure tensor.

In the phase space of the sub-system,
the zeroth and first momentum moments are
\begin{equation}
{\bf P}_{0}({\bf \Gamma}) = \sum_{i=1}^N  {\bf p}_{i}
, \mbox{ and }
\underline{ \underline {\mathrm P}}_1({\bf \Gamma}) =
\sum_{i=1}^N {\bf q}_i {\bf p}_{i} .
\end{equation}
The zeroth moment is a conserved variable,
and hence its adiabatic rate of change is zero.
The adiabatic rate of change of the first momentum moment is
\begin{eqnarray} \label{Eq:dotPx10}
%\lefteqn{
\dot{\underline{ \underline {\mathrm P}}}^0_1({\bf \Gamma})
%} \nonumber \\
& = &
%\dot{\bf \Gamma}^0 \cdot \nabla P_{x1}({\bf \Gamma})
%\nonumber \\ & = &
\sum_{i=1}^N  \left[ \dot{\bf q}_i^0 \cdot \nabla_{{q}_i}
 \underline{ \underline {\mathrm P}}_1({\bf \Gamma})
+
\dot{\bf p}_i^0 \cdot  \nabla_{{ p}_i}
\underline{ \underline {\mathrm P}}_1({\bf \Gamma})
\right]
\nonumber \\ & = &
\sum_{i=1}^N
\left[  \dot{\bf q}_i^0  {\bf p}_i
+ \dot {\bf p}_i^0 {\bf q}_i  \right]
%\nonumber \\ & = &
%\sum_{i=1}^N \left[ \frac{ {p}_{zi} p_{xi} }{m} -
%z_i \sum_{j=1}^N \!\!^{(j \ne i)}
%u'(q_{ij}) \frac{x_{i}-x_{j}}{q_{ij}} \right]
%\nonumber \\ & = &
%\sum_{i=1}^N   \frac{ {p}_{zi} p_{xi} }{m}
%- \sum_{i<j}^N u'(q_{ij}) \frac{[z_i-z_j] [x_{i}-x_{j}]}{q_{ij}}
\nonumber \\ & = &
\sum_{i=1}^N \left[ \frac{ {\bf p}_i {\bf p}_i }{m}
+ {\bf F}_i {\bf q}_i  \right].
\end{eqnarray}
Defining the peculiar momentum as
$\tilde{\bf p}_i \equiv {\bf p}_i - m {\bf v}({\bf q}_i)$,
this may be written as
\begin{equation}
\underline{\underline{ \dot {\mathrm P}_{1}^0}}({\bf \Gamma})
=
\sum_{i=1}^N
\left[ \frac{ \tilde{\bf p}_i \tilde{\bf p}_i }{m}
+ {\bf F}_i {\bf q}_i  \right]
+ m \sum_{i=1}^N
{\bf v}({\bf q}_{i}) {\bf v}({\bf q}_{i}) .
\end{equation}
Removing the convective  momentum flux as discussed above,
the pressure tensor is therefore
\begin{eqnarray}
\underline{\underline {\mathrm P}}({\bf \Gamma})
& = &
\frac{1}{V}
\underline{\underline{ \dot {\mathrm P}_{1}^0}}({\bf \Gamma})
- \frac{m}{V}  \sum_{i=1}^N
{\bf v}({\bf q}_{i}) {\bf v}({\bf q}_{i})
\nonumber \\ &=&
\frac{1}{V}
\sum_{i=1}^N \left[ \frac{ \tilde {\bf p}_{i} \tilde {\bf p}_{i} }{m}
+ {\bf F}_{i} {\bf q}_i \right]
\nonumber \\ & = &
\frac{1}{V}\sum_{i=1}^N \frac{ \tilde {\bf p}_{i} \tilde {\bf p}_{i} }{m}
- \frac{1}{V} \sum_{i<j}^N u'(q_{ij}) \frac{{\bf q}_{ij} {\bf q}_{ij}}{q_{ij}}.
\end{eqnarray}
The final equality holds for a pair-wise additive potential
and it shows explicitly the symmetry of the pressure tensor.
The Irving-Kirkwood stress tensor is the negative of this.

The linear constitutive relation from hydrodynamics
relates the most likely value of the traceless part
of the viscous pressure tensor to
the traceless symmetric part of the velocity gradient tensor,
\begin{equation}
\overline{{ \underline{\underline \Pi}}}^*
=
- 2\eta  \left[ \nabla {\bf v}({\bf r}) \right]^{*,\mathrm{sym}}
\equiv - 2\eta \underline{\underline v}_1^{*,\mathrm{sym}},
\end{equation}
where $\eta$ is the shear viscosity.
In component form this is
\begin{equation}
\overline{\Pi}^*_{\alpha\beta}
=
- \eta  \left[\frac{\partial v_\alpha({\bf r})}{\partial r_\beta}
+ \frac{\partial v_\beta({\bf r})}{\partial r_\alpha}
-\frac{2}{3} \delta_{\alpha\beta} \nabla \cdot {\bf v}({\bf r})
\right] .
\end{equation}
This result holds in a most likely sense,
going forward in time.\cite{NETDSM}
%Recall that the first velocity is the velocity gradient.
%which most likely equals  the applied shear rate,
%$\overline v_{sx1}(z) = \gamma$.
%This equality holds even in unsteady turbulent flow,
%since at any point
%the unsteady part of the flow averages to zero.
%(Actually this result holds for a uniform system,
%or else when averaged over the whole system.
%As will be discussed in the results section,
%the structure of the fluid can be different
%in a slab region next to the boundary where
%the fluctuation-dissipation parameter is applied.
%Hence the viscosity and the shear rate can also differ in this region.)

%%%%%%%%%%%%%%%%%%%%%%%%%%%
\subsubsection{Reservoirs} \label{Sec:Res}

Consider two reservoirs
beyond  boundaries  located at $z=\pm L_z/2$,
moving with velocities  $v_{\mathrm{r}x\pm} =\pm v_{\mathrm{r}x}$.
The zeroth and first velocities exerted by the reservoirs
on the sub-system  are\cite{NETDSM}
\begin{eqnarray}
v_{\mathrm{r}x0} & \equiv &
\frac{1}{2} \left[  v_{\mathrm{r}x+} + v_{\mathrm{r}x-} \right]
\nonumber \\
\mbox{and }
\gamma \equiv v_{\mathrm{r}x1} & \equiv &
\frac{1}{L_z} \left[  v_{\mathrm{r}x+}  - v_{\mathrm{r}x-} \right] .
\end{eqnarray}
The zeroth velocity is the average or mid velocity,
and the first velocity is the velocity gradient or applied shear rate $\gamma$.
It is emphasized that the applied shear rate $\gamma$
(as well as the applied temperature $T$ that enters below)
is a property of the reservoirs.
As such it is independent of the state of the sub-system,
which means that
it has the same value if the resultant sub-system motion is steady shear flow
or if it undergoes unsteady or turbulent flow.
This observation means
that using $T$ or $\gamma$ in the present equations of motion
does not assume a particular sub-system profile,
nor does it introduce any unintended bias into those equations of motion.

The phase space probability distribution
is essentially the exponential of the reservoir entropy,
$ S_\mathrm{r}({\bf \Gamma},t)
= S_\mathrm{r,st}({\bf \Gamma}) + S_\mathrm{r,dyn}({\bf \Gamma},t) $.
Since the reservoir velocities are equal and opposite,
$v_{\mathrm{r}x0} = 0$
and $v_{\mathrm{r}x1} = 2v_{\mathrm{r}x+}/L_z \equiv \gamma$.
In this case,
since energy and momentum are exchangeable with the reservoir,
and since velocity is conjugate to momentum,
the  static part of the reservoir entropy is
\begin{eqnarray} \label{Eq:Srst-shear}
S_\mathrm{r,st}({{\bf \Gamma}})
& = &
\frac{-{\cal H}({\bf \Gamma})}{T}
+ \frac{\gamma}{T}  P_{zx1}({\bf \Gamma}) .
%\nonumber \\  & = &
%\frac{-1}{T} \sum_{i\alpha} \frac{ \left[ p_{i\alpha}
%- m_i \gamma z_i  \delta_{x\alpha} \right]^2 }{2m_i}
%\nonumber \\ && \mbox{ }
%+ \frac{\gamma^2 }{2T} \sum_{i} m_i z_i^2
%-\frac{1}{T} \sum_{i<j}^N u(q_{ij}) .
\end{eqnarray}
The first term on the right-hand side %of the first equality
arises from the exchange of energy with the reservoirs,
and the second term from the exchange of momentum.
As mentioned, both the temperature $T$ and the shear rate $\gamma$
that appear here are properties of the reservoir
that are independent of the state of the sub-system.
Likewise the product $\gamma z_i$ is dictated by the spatial
arrangement of the reservoirs
and one does not have the thermodynamic freedom to replace this by
some local quantity that depends on the state of the sub-system.
%The first term on the right-hand side of the rearrangement
%that is the second equality
%is the peculiar kinetic energy,
%which is the kinetic energy in the frame of reference moving
%with the nominal local flow.
%The rearrangement is mathematically exact
%and it makes no statement about the local physical state of the sub-system.
%The second term in the second equality is second-order in the shear rate
%and it would create a spatial inhomogeneity
%(ie.\ a density depletion in the center of the sub-system)
%at high shear rates
%if it were not canceled by the part quadratic in $\gamma$
%in the peculiar kinetic energy.

With this expression for the static part of the reservoir entropy
for shear flow,
the components of the unconstrained dissipative force are
\begin{eqnarray}
\overline R_{i\alpha}^\mathrm{f}({\bf \Gamma})
& = &
\frac{ \theta_{i\alpha} |\Delta_t| }{2k_\mathrm{B}}
\frac{\partial  S_\mathrm{r,st}({\bf \Gamma})}{\partial p_{i\alpha}}
+ \frac{\theta_{i\alpha} |\Delta_t| }{2k_\mathrm{B}}
\left( \widehat t  - 1 \right)
 \overline{S'_{\mathrm{p}i\alpha}}
%\nonumber \\ & = &
%\frac{ -\theta_{i\alpha} |\Delta_t| }{2k_\mathrm{B}}
%\left[ \frac{1}{T}
%\frac{\partial {\cal H}({\bf \Gamma})}{\partial p_{i\alpha}}
%-
%\frac{\gamma}{T}
%\frac{\partial P_{x1}({\bf \Gamma})}{\partial p_{i\alpha}}
%\right]
%\nonumber \\ & & \mbox{ }
%+ \frac{\theta_{i\alpha} |\Delta_t| }{2k_\mathrm{B}}
%\left( \widehat t  - 1 \right)
%\overline{
%\frac{\partial S_\mathrm{r,st}({\bf \Gamma})}{\partial p_{i\alpha}} }
\nonumber \\ & = &
\frac{ -\theta_{i\alpha} |\Delta_t| }{2k_\mathrm{B}}
\left[
\frac{ p_{i\alpha} }{ T m_i }
-
\frac{ \gamma}{ T  }  z_i \delta_{\alpha x}
\right]
\nonumber \\ & & \mbox{ }
- \frac{\theta_{i\alpha} |\Delta_t| }{2k_\mathrm{B}}
\left( \widehat t  - 1 \right)
\left[
\frac{ \overline p_{i\alpha} }{ T m_i }
-
\frac{ \gamma }{ T  }  \overline z_i   \delta_{\alpha x}
\right]
\nonumber \\ & = &
\frac{ -\theta_{i\alpha} |\Delta_t| }{2 m_ik_\mathrm{B}T}
\left[ p_{i\alpha} - m_i \gamma z_i \delta_{\alpha x} \right]  .
\end{eqnarray}
%Recall that the first reservoir velocity is the applied velocity gradient
%(equivalently, the applied shear rate).
One can recognize  the bracketed term as the peculiar momentum,
which is on average zero (to ${\cal O}(\gamma^3)$).
The  constant term in the first and second equalities
vanishes for a forward time step,
since $ \widehat t  \equiv \mbox{sign}(\Delta_t) = 1$.
The final result is of the same form as a hydrodynamic drag force
in the  Langevin equation,
but using the local fluctuating velocity of the atom.\cite{Kremer90,Shang17}
Unlike  the Langevin equation and its peculiar extension,
the present result has a rigorous
thermodynamic and statistical mechanical justification.

%%%%%%%%%%%%%%%%%%%%%%%%%%%
\subsubsection{Constraint for Shear Flow}

As was discussed above
it is useful to add a constraint
to the dissipative reservoir force
to make it orthogonal to the gradient of the adiabatic rate of change
of the static part of the reservoir entropy.
In the case of shear flow,
the latter is proportional to
$\dot P_{zx1}^0$.
Hence the constraint to be applied is
\begin{equation} \label{Eq:constraint}
\overline {\bf R}({\bf \Gamma})
\cdot \nabla_\mathrm{p} \dot P_{zx1}^0({\bf \Gamma})  = 0 .
\end{equation}
Using the expression given above,
one has explicitly
\begin{equation}
\frac{\partial \dot P_{zx1}^0({\bf \Gamma}) }{\partial p_{i\alpha}}
=
m_i^{-1} \left[ p_{xi} \delta_{z\alpha} + p_{zi} \delta_{x\alpha} \right] .
\end{equation}

With the unconstrained dissipative force $\overline {R}_{i\alpha}^\mathrm{f}$
%being the $i\alpha$ component of the unconstrained dissipative force
given above,
one can introduce a Lagrange multiplier $\mu$
so that the constrained dissipative force is
\begin{equation}
\overline {R}_{i\alpha}
=
\overline {R}_{i\alpha}^\mathrm{f}
+ \mu \theta_{i\alpha}
m_i^{-1} \left[ p_{xi} \delta_{z\alpha} + p_{zi} \delta_{x\alpha} \right] .
\end{equation}

The Lagrange multiplier is given by
\begin{equation}
\mu({\bf \Gamma},t)
=
\frac{-\sum_{i,\alpha}
\overline {R}_{i\alpha}^\mathrm{f}
m_i^{-1} \left[ p_{xi} \delta_{z\alpha} + p_{zi} \delta_{x\alpha} \right]
}{
\sum_{i,\alpha} \theta_{i\alpha}
m_i^{-2} \left[ p_{xi} \delta_{z\alpha} + p_{zi} \delta_{x\alpha} \right]^2
}.
\end{equation}
Obviously both the free $\overline {R}_{i\alpha}^\mathrm{f}$
and the constrained $\overline {R}_{i\alpha}$ dissipative force,
as well as the stochastic force $\tilde {R}_{i\alpha}$,
vanish for molecules in regions where
$\theta_{i\alpha} \equiv \theta(z_i)= 0$.

%%%%%%%%%%%%%%%%%%%%%%%%%%%
\subsubsection{Equipartition Theorem for Shear Flow}

For shear flow, the equipartition  theorem, Eq.~(\ref{Eq:EquiThm}),
with the above result for the static part of the reservoir entropy,
Eq.~(\ref{Eq:Srst-shear}),
may be rearranged as
\begin{equation} \label{Eq:EquiThm-Shear}
3 N m k_\mathrm{B}T
=
\sum_{i\alpha}
\left[ p_{i\alpha}
- m \gamma z_i  \delta_{x\alpha} \right]^2
+ {\cal O}(\gamma^2).
\end{equation}
The  terms that appear explicitly here may be recognized
as being the same as for the equilibrium equipartition theorem,
but for the peculiar kinetic energy rather than the total kinetic energy.
One can use this to define a local kinetic temperature,
$T_\alpha^\mathrm{kin}(z)$,
but this is not equal to the actual sub-system temperature
except at low shear rates.

%%%%%%%%%%%%%%%%%%%%%%%%%%%
\subsection{Viscous Heating}

In shear flow, viscous heating raises the temperature of the sub-system
until it is balanced by heat conduction to the boundary.
The rate of doing work is the velocity times the force,
and the latter is the rate of change of momentum.
The rate of work performed on a slab $A \mathrm{d}z$ at $z=0$ is
\begin{eqnarray}
\dot W
& = &
 - A  \left[ \overline{ \Pi}_{xz} \overline v_x(\mathrm{d}z/2)
-  \overline{ \Pi}_{xz} \overline v_x(-\mathrm{d}z/2) \right]
\nonumber \\ & = &
 A \mathrm{d}z \,  \eta \gamma^2  .
\end{eqnarray}
The negative sign is because $\overline{ \Pi}_{xz}$ is
the momentum leaving the slab through the upper face.
%The rate of doing work is positive and quadratic in the shear rate,
%as it must be.
%Since the dimensions of viscosity are $[\eta] = M L^{-1} s^{-1}$,
%the right hand side has dimensions $L^3 M L^{-1} s^{-1} s^{-2} = E  s^{-1}  $,
%which is correct.
With this,
the rate of change of temperature is
\begin{eqnarray}
\dot T(z)
& = &
c_p^{-1} \left[ \dot w(z)
- \frac{1}{A} \nabla \cdot \overline {\bf J}_\mathrm{E}^0(z)
\right]
\nonumber \\ & = &
c_p^{-1} \left[  \eta \gamma^2 + \lambda  \nabla^2 T(z) \right] ,
\end{eqnarray}
where $c_p$ is the heat capacity per unit volume
and $\lambda$ is the thermal conductivity.
In a steady state system the temperature is constant,
$\dot T(z)=0$,
and so
\begin{eqnarray} \label{Eq:T(z)}
T(z)  & = &
\frac{-\eta  \gamma^2}{2\lambda} z^2 + \mbox{const.}
\nonumber \\ & = &
 T
 +
\frac{\eta  \gamma^2}{2\lambda} \left[ \frac{L_z^2}{4} - z^2 \right].
\end{eqnarray}
The second equality holds if the temperature at the boundary
equals that of the reservoir.
This expression is valid in regions
where the heat being extracted by a thermostat is negligible
(ie.\ regions of adiabatic evolution).
Similar expressions have been obtained by Khare \emph{et al.}\cite{Khare97}
and by Hoover {\emph{et al.}\cite{Hoover08}
%and is confirmed by the simulation results reported below.

This variation in temperature
is obviously significant
for large shear rates and large systems.
This is one source of non-Newtonian behavior,
and, depending on the sign of $\eta' \equiv \partial \eta/\partial T$
(both signs occur in the results below),
it will contribute to thickening or thinning with increasing shear rate.
This also causes the viscosity to depend on the system size
at high shear rates,
which is extremely unusual in thermodynamics,
since, at least in the equilibrium case,
in the thermodynamic limit
all parameters are either intensive
(ie.\ independent of system size)
or else extensive (ie.\ linearly proportional to the system size).

%For example, if one averages the temperature change
%over the center half of the system,
%$\Delta_T(\gamma,L_z) = ( \eta  \gamma^2/2\lambda)(11 L_z^2/48)$,
%then $\eta(\gamma,L_z) = \eta(\gamma) + \Delta_T(\gamma,L_z) \eta'$,
%where $\eta' = \partial \eta/\partial T$.

This temperature dependence and non-intensive behavior poses
conceptual challenges in how to apply
results for the shear viscosity.
It is emphasized that this is a real physical effect,
since in steady state shear heat can only be dissipated via the boundaries.

One pragmatic solution is to perform simulations in the regime
where the temperature increase is relatively negligible,
$T(0)-T \ll T$, or
\begin{equation} \label{Eq:low-gamma-lim}
%\frac{\eta  \gamma^2 L_z^2}{8\lambda T} \ll 1,
%\mbox{ or }
\gamma \ll \sqrt{ \frac{8\lambda T}{\eta   L_z^2} }
\mbox{ or }
|v_\mathrm{r,+} - v_\mathrm{r,-}| \ll \sqrt{ \frac{8\lambda T}{\eta } }.
\end{equation}
If this holds, we may assume that the sub-system temperature
is the same as the reservoir temperature.
In this case the shear viscosity can be expected to be independent
of system size, $\eta(\gamma,T,\rho,L_z) \Rightarrow \eta(\gamma,T,\rho)$.
In this regime the temperature of the sub-system
is the same as the temperature of the reservoir,
and there is no ambiguity as to which temperature
is referred to.

In the real world or laboratory situation,
very high shear rates only occur over a limited spatial extent,
such as in a shock wave,
or in the vicinity of an injection aperture,
or around a sharp protuberance.
In addition, very high shear is often of a transient nature,
and the amount of viscous heat produced may be insignificant
compared to the heat capacity of the relevant volume.
In these cases one can assume that the temperature
in the shearing region is constant,
and one can use the results for the shear viscosity
that are obtained in the limit where it is independent of the system size,
and where the temperature of the sub-system is approximately uniform
and equal to that of the reservoir.

%%%%%%%%%%%%%%%%%%%%%%%%%%%
\subsection{Pressure}

The normal component of the thermodynamic pressure is now analyzed.
With the applied shear rate being
$\gamma = [v_{{\mathrm r}+}-v_{{\mathrm r}-}]/L_z$,
the $L_z$ derivative can be made at constant shear rate $\gamma$
or else at constant reservoir velocities $v_{{\mathrm r}\pm}$.
The second of these is arguably the case with greatest physical relevance.
One can define the $z$-pressure
as the derivative of the total unconstrained entropy,
\begin{eqnarray}
p_z & \equiv &
\frac{T}{A} \frac{\partial S_\mathrm{tot}(N,V,T,\gamma)}{\partial L_z}
\nonumber \\ & = &
\frac{k_\mathrm{B}T}{A Z_\mathrm{tot}}
\frac{\partial  }{\partial L_z}
\int  \mathrm{d}{\bf \Gamma} \;
e^{S_\mathrm{r}({\bf \Gamma})/k_\mathrm{B}}
\nonumber \\ & = &
\frac{k_\mathrm{B}T}{A Z_\mathrm{tot}}
\left\{\frac{ 3 N}{ L_z } Z_\mathrm{tot}
+
\int  \mathrm{d}{\bf \Gamma} \;
e^{S_\mathrm{r}({\bf \Gamma})/k_\mathrm{B}}
\frac{\partial S_\mathrm{r}({\bf \Gamma})}{k_\mathrm{B}\partial L_z}
\right\}
\nonumber \\ & = &
\frac{3Nk_\mathrm{B}T}{V}
+
\frac{T}{A}  \left<
\frac{\partial S_\mathrm{r}({\bf \Gamma})}{\partial L_z}
\right>_\mathrm{ne} .
\end{eqnarray}
The first term is that of an ideal gas,
and it follows from the usual re-scaling of the sub-system.
As mentioned above the reservoir entropy is the sum of
static and dynamic parts,
with the static part being given by Eq.~(\ref{Eq:Srst-shear}),
$ S_\mathrm{r,st}({{\bf \Gamma}})
=
-{\cal H}({\bf \Gamma})/T
+ \gamma   P_{zx1}({\bf \Gamma})/T $.
Writing $q_{iz} = L_z \tilde q_{iz}$,
and supposing that the potential energy
is the sum of pair potentials,
one has
\begin{eqnarray}
\frac{\partial {\cal H}({\bf \Gamma})}{\partial L_z}
& = &
\sum_{i<j} \frac{\partial u(q_{ij}) }{\partial L_z}
\nonumber \\ & = &
\frac{1}{L_z}
\sum_{i<j} u'(q_{ij}) \frac{z_{ij}^2}{q_{ij}}.
\end{eqnarray}
Also, with the first momentum moment being
$ P_{zx1}({\bf \Gamma}) = \sum_{i=1}^N z_i p_{xi}$,
a similar scaling yields
\begin{equation}
\frac{\partial  P_{zx1}({\bf \Gamma}) }{\partial L_z}
=
\left\{ \begin{array}{ll}
L_z^{-1} P_{zx1}({\bf \Gamma}), & \gamma = \mbox{const.},\\
0, & v_{{\mathrm r},\pm} = \mbox{const.}\\
\end{array} \right.
\end{equation}
Hence one has (for the case $\gamma=\mbox{const.}$)
\begin{eqnarray}
p_z & = &
\frac{3Nk_\mathrm{B}T}{V}
- \frac{1}{V}
 \left< \sum_{i<j} u'(q_{ij}) \frac{z_{ij}^2}{q_{ij}}
\right>_\mathrm{ne}
 \\ & & \mbox{ }
+ \frac{\gamma}{V}  \left<  P_{zx1}({\bf \Gamma}) \right>_\mathrm{ne}
+
\frac{T}{A}  \left<
\frac{\partial S_\mathrm{r,dyn}({\bf \Gamma})}{\partial L_z}
\right>_\mathrm{ne} .\nonumber
\end{eqnarray}
The first two terms may be recognized as essentially
the non-equilibrium average of
the usual equilibrium virial expression for the pressure.
Since $\overline p_{xi} = m \gamma z_i + {\cal O}(\gamma^3)$,
the third term for the case $\gamma=\mbox{const.}$ is
\begin{eqnarray}
\frac{\gamma}{V}  \left<  P_{zx1}({\bf \Gamma}) \right>_\mathrm{ne}
& = &
\frac{\gamma}{V} m \gamma \sum_{i=1}^N \langle z_i^2 \rangle_\mathrm{ne}
+ {\cal O}(\gamma^4)
\nonumber \\ & = &
\frac{m n \gamma^2  L_z^2}{12} + {\cal O}(\gamma^4).
\end{eqnarray}
The final approximation assumes a uniform density profile.
%The simulations reported below show the final analytic form to be surprisingly
%accurate over the entire range of shear rates studied.
One sees that there is a repulsive contribution to the pressure
that goes as $\gamma^2 L_z^2$.
This contrasts with a uniform equilibrium system
where the pressure is intensive,
which is to say that it is independent of $L_z$.
This situation is similar to that of an inhomogeneous equilibrium  fluid
(since a shearing system is also inhomogeneous, Eq.~(\ref{Eq:T(z)})),
such as a slit pore,
where the normal component of the pressure also depends
on separation $L_z$.
In a sense it is worse in the present case in that
the pressure actually diverges quadratically with sub-system width.

As mentioned this derivative at constant shear rate $\gamma$
is different to the derivative at constant reservoir velocity
$v_\mathrm{r\pm}$.
In the latter case $ L_z\gamma$ is constant
and the $L_z$ derivative of the first momentum moment vanishes.
In this arguably more physical case,
the normal component of the pressure
does not have such a divergent non-extensive contribution.
It likely still has an $L_z$ dependence due to the inhomogeneous nature
of the shearing system.

% end comment %%%%%%%%%%%%%%%%%%%%%%%%%%%%%%%%%%%%%%%%%

%%%%%%%%%%%%%%%%%%%%%%%%%%%
\subsection{SDMD Simulation Details} \label{Sec:SimDet}

A Lennard-Jones fluid was simulated,
with cut and shifted pair potential
\begin{equation}
u(r) =
\left\{ \begin{array}{ll}
\displaystyle
4 \varepsilon
\left[ \frac{\sigma^{12}}{r^{12}} -\frac{\sigma^6}{r^6} \right]
- u_\mathrm{cut}, &
r \le R_\mathrm{cut}, \\
0, & r > R_\mathrm{cut} .
\end{array}
\right.
\end{equation}
Here $u_\mathrm{cut} =
4 \varepsilon \left[ (\sigma/R_\mathrm{cut})^{12}
- (\sigma/R_\mathrm{cut})^6  \right]$,
but in fact this is immaterial.
All of the following results are made dimensionless
using the  well-depth energy $\varepsilon$,
the atomic diameter $\sigma$,
and the mass $m$.
These give the unit of time as
$ \sqrt{ m \sigma^2 /\varepsilon }$,
which is 2.2\,ps for argon.

The cut-off radius was $R_\mathrm{cut} = 2.5$.
It can be shown that the consequently neglected tail contribution
to the shear viscosity
is ${\cal O}(R_\mathrm{cut}^{-9})$, which is negligible.
This was confirmed numerically by performing one simulation with
$R_\mathrm{cut} = 3.5$,
where it was found that the shear viscosity changed
by 0.09\%,
which is insignificant
compared to the 0.4\% statistical standard  error for those runs.

A rectangular simulation cell was used,
usually with $L_x = L_y \approx 0.9 L_z$.
In all cases 1000 atoms were used.
Periodic boundary conditions were applied,
with Lees-Edwards sliding brick conditions imposed
in the $z$-direction.\cite{Lees72}

A small cell spatially based neighbor table was used,
with cells cubic and of edge length 0.7.\cite{AttardV,AttardIX}
This typically gave a  neighborhood volume 2.6 times
the cut-off sphere, which is much better
than typical large cell neighbor tables achieve.
The neighbor table was modified for the Lees-Edwards conditions
by adding a separate neighbor table of width $R_\mathrm{cut}$
beyond the upper and lower boundaries
and maintaining a record of the images of the atoms in this region.
This extension to the neighbor table was fixed to the central neighbor table,
and the image atoms moved through it
in accord with the Lees-Edwards conditions.
The height of the central cell $L_z$ was necessarily an integer multiple
of the small cell edge length,
but not necessarily so for the extended system
$L_z+2R_\mathrm{cut}$.
The neighbor table performed remarkably well,
consistent with previous experience
where the simulations scaled linearly with the number of atoms.
\cite{AttardV,AttardIX}

The Verlet leap frog algorithm was used.\cite{Shang17,Gao16}
This is second order in the time step,
and is about an order of magnitude more efficient
than simple time stepping.
The time step was $\Delta_t = 0.01$,
which was halved in some tests.
Higher order methods were not explored.\cite{Shang17}
%It should be mentioned that derivation
%of the stochastic, dissipative equations of motion
%from the non-equilibrium transition probability,
%and the proof that they preserve the non-equilibrium
%probability density in phase space, has only been carried out
%to linear order in the time step.\cite{NETDSM,STD2}
%The question of the appropriateness
%of using a second order method to solve these equations
%has not been addressed here.

Usually simulations were performed for
$10^6$-- $10^7$ time steps,
with high shear rates requiring fewer time steps.
To obtain the temperature derivative of the viscosity
$ 4 \times 10^7$ time steps were used in total
at each state point.
Each case was usually started from a nearby shear rate,
after velocity re-scaling and position pinching,
and a brief equilibration period.
The neighbor table was updated every time step.
Values for averaging were accumulated once every 20 time steps.
In hindsight, it would be more efficient to accumulate these
more frequently.
Each simulation was broken into 50 blocks,
the standard deviation of which were estimated
from the fluctuations between them.
The statistical error quoted everywhere below
is one standard error on the mean,
which is this standard deviation divided by $\surd 50$.
It gives the 68\% confidence interval.
%It would be better to use $\pm 1.96 \sigma/\surd 50$,
%as this gives the 95\% confidence level
%(probability that this method would estimate the true value
%as lying within this interval 95% of the time.)
%%  deviation $\sigma $  or the variance $\sigma^2$.
%%About 68\% of values drawn from a normal distribution
%%are within one standard deviation $\sigma $ away from the mean;
%%about 95\% of the values lie within two standard deviations;
%%and about 99.7\% are within three standard deviations.

%%%%%%%%%%%%%%%%%%%%%%%%%%%%%%%%%%%%%%%%%%%%%%%%%%%%%%%%%%%%%%%%%%%%%%%%%%
%
\section{Results}
\setcounter{equation}{0} \setcounter{subsubsection}{0}
%
%%%%%%%%%%%%%%%%%%%%%%%%%%%%%%%%%%%%%%%%%%%%%%%%%%%%%%%%%%%%%%%%%%%%%%%%%%

%%%%%%%%%%%%%%%%%%%%%%%%%%%%%%%%%%%%%%%%%%%%%%%%%%%%%%%%%%%%%%%%%%
\begin{figure}[t!]
\centerline{
\resizebox{8cm}{!}{ \includegraphics*{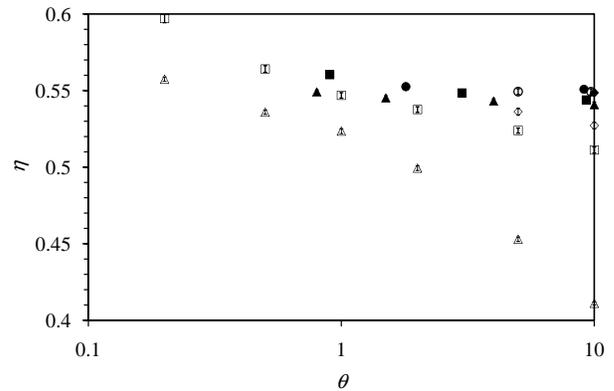} } }
\caption{\label{Fig:eta-theta}
Shear viscosity versus fluctuation-dissipation parameter
at $T=2$, $\rho=0.452$, and $\gamma=0.1$.
The fluctuation-dissipation form is
$n=0$ (triangles), $n=1$ (squares), $n=2$ (diamonds),
and boundary slab (circles).
The empty symbols are unconstrained,
and the filled  symbols are constrained.
The error bars are obscured by the symbols here and throughout.
}
\end{figure}
%%%%%%%%%%%%%%%%%%%%%%%%%%%%%%%%%%%%%%%%%%%%%%%%%%%%%%%%%%%%%%%%%%

Figure~\ref{Fig:eta-theta}
shows the viscosity as a function of the fluctuation-dissipation parameter.
The state point was $T=2$, $\rho=0.452$,
and for these simulations, $N=1000$, $L_x=L_y=12.90$, and $L_z=13.30$.
The shear rate is fixed at $\gamma=0.1$, which corresponds to
45\,GHz for argon.
Based on results below at this state point,
this can be considered to be in the linear, low shear rate regime.
The  limit for negligible temperature change, Eq.~(\ref{Eq:low-gamma-lim}),
is $\gamma_\mathrm{low} = 0.61$,
using  $\lambda=2.25$\cite{AttardV} and $\eta=0.55$.

First focussing on the uniform application
of the  fluctuation-dissipation parameter, $n=0$,
and the unconstrained equations of motion
(empty triangles),
one sees that the simulated viscosity
is strongly affected by the strength of the parameter,
with the viscosity decreasing as the parameter is increased.
It will be recalled that the stochastic, dissipative equations of motion
are correct to linear order in $\theta$,
which means that they are unreliable for large values.
Enforcing the constraint, Eq.~(\ref{Eq:constraint}) (filled triangles),
largely removes this sensitivity.

The influence of $\theta$
on the viscosity is also reduced when it
is focussed in the boundary region,
$n=1$ (squares), $n=2$ (diamonds),
and confined to the boundary slab $|z| > L_z/2-R_\mathrm{cut}$  (circles).
In these cases the viscosity (and the temperature)
are measured in the central half of the system, $|z| < L_z/4$
here and below.
In the slab case the evolution in this central region is strictly adiabatic.
One can see that at $\theta=10$, the unconstrained results (open symbols)
increase toward the constrained, uniform result
as the dissipative and stochastic forces
are more narrowly confined to the boundary.
There is relatively good agreement between the different boundary methods.
The slight disagreement
between the constrained and unconstrained boundary slab results
at $\theta =10$ indicates that the structure and motion
in the boundary slab region
can still influence the viscosity measured in the central region.

If the fluctuation-dissipation parameter is
too small to remove the viscous heat,
then the sub-system kinetic temperature will
be larger than the reservoir temperature,
particularly for a boundary-confined parameter.
This will effect the viscosity  even in the constrained case.
For example,
for the $n=2$ constrained case,  at $\theta=0.5$,
the central viscosity was $\eta = 0.5842 \pm 0.0030$
and the central kinetic temperature was $T=2.2079 \pm 0.0019$.
The  temperature derivative at this state point was measured to be
$\partial \eta/\partial T = 0.0800 \pm 0.0089$
(measured at $\gamma=0.1$ with $\theta=10$ and $n=0$).
Using this to correct the measured viscosity to $T=2$ gives
$\eta = 0.5668  \pm 0.0045$,
which is closer to the values measured at $\theta =10$,
$\eta=0.55$.
The correction is not exact because of additional effects
from the induced inhomogeneity in both temperature and density.

%(since here $\partial \eta /\partial T > 0$).
%Focussing now on the constrained equations of motion (filled symbols),
%Fig.~\ref{Fig:eta-theta} shows that the viscosity is relatively insensitive
%to the value of the fluctuation-dissipation parameter
%for both the boundary-focussed and the uniform forms.
%The small increase with decreasing parameter
%appears mainly due to the temperature increase
%from viscous heating when  the dissipative force
%lacks the strength to remove heat fast enough at the reservoir temperature.

This particular state point, $T=2$, $\rho=0.452$,
was chosen to make quantitative comparison
with the viscosity simulated with four different methods
by Hess.\cite{Hess02}
Hess obtained
from the transverse current method
$\eta = 0.433 $ and $\eta = 0.465 $,
from the pressure fluctuations (Green-Kubo) method
$\eta = 0.444 \pm 0.018$,
from a periodic external perturbation method
$\eta = 0.443 \pm 0.002$,
and from the NEMD (SLLOD, $\gamma=0.5$) method $\eta = 0.462 \pm 0.008$.
These are in agreement with each other but they are significantly smaller
than the present best estimate of the shear viscosity,
$\eta=$0.54--0.55.
However, the four  types of simulation performed by Hess
are not independent of each other
as they were all deterministic molecular dynamics
using a uniform Berendsen thermostat with
the same coupling time of 20 (all methods).\cite{Hess02}
Since this is presumably  large enough to hold the kinetic temperature
at the specified reservoir value,
the most direct comparison with the present results
is the uniform, $n=0$, unconstrained case,
with $\theta=5$ being similarly large enough.
This case gives $\eta=0.4530 \pm 0.0015$,
which is comparable to Hess' estimates.
(As mentioned in \S \ref{Sec:SimDet} above,
the cut-off radius,  $R_\mathrm{cut}=2.5$ here,
$R_\mathrm{cut}=5$ in Ref.~[\onlinecite{Hess02}],
has negligible effect on the viscosity.)

% end comment %%%%%%%%%%%%%%%%%%%%%%%%%%%%%%%%%%%%%%

%%%%%%%%%%%%%%%%%%%%%%%%%%%%%%%%%%%%%%%%%%%%%%%%%%%%%%%%%%%%%%%%%%
\begin{figure}[t!]
\centerline{
\resizebox{8cm}{!}{ \includegraphics*{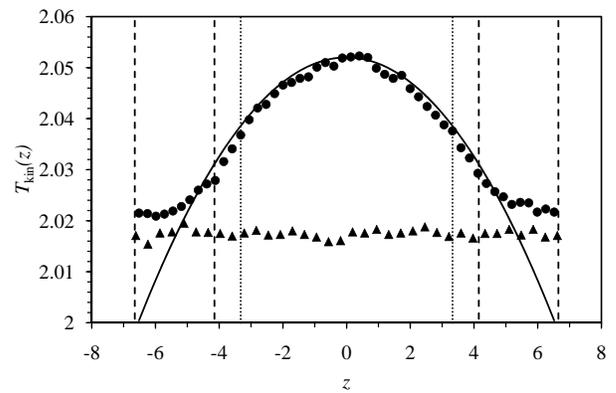} } }
\caption{\label{Fig:T(z)}
Kinetic temperature profile at $T=2$, $\rho=0.452$,
and $\gamma=0.1$.
The constrained equations of motion are used
with $\theta=10$, $n=0$ (triangles),
and $\theta=10$, boundary slab (circles).
The solid curve is the expected adiabatic temperature profile,
Eq.~(\ref{Eq:T(z)}) using $\eta=0.546$ and $\lambda=2.25$,\cite{AttardV}
with $T(0)$ fitted.
The dashed lines delineate the boundary slab,
and the dotted lines delineate the central half of the sub-system.
The system is periodic beyond the plotted data.
}
\end{figure}
%%%%%%%%%%%%%%%%%%%%%%%%%%%%%%%%%%%%%%%%%%%%%%%%%%%%%%%%%%%%%%%%%%

Figure~\ref{Fig:T(z)} shows the local kinetic temperature,
 Eq.~(\ref{Eq:EquiThm-Shear}).
It can be seen that the uniform  fluctuation-dissipation parameter,
$\theta=10$, $n=0$,
is relatively good at maintaining a uniform kinetic temperature,
with average value $T_\mathrm{kin}=2.01747 \pm 0.00013$.
The difference from the reservoir temperature $T=2$
appears to correspond to the neglected $\gamma^2$ corrections
to the kinetic temperature.
For the boundary slab case,
the kinetic temperature profile is parabolic in the central adiabatic region,
with the curvature given by  Eq.~(\ref{Eq:T(z)}),
using the thermal conductivity $\lambda=2.25$
(interpolated from Ref.~[\onlinecite{AttardV}])
and the present viscosity $\eta=0.546$.
The kinetic temperature profile (circles) has slightly larger curvature
than the macroscopic temperature profile (curve),
which may either be interpreted as a small change
in the value of the shear viscosity or the thermal conductivity,
possibly because of the inhomogeneities,
or else as weak higher order contributions to the equipartition theorem.

%%%%%%%%%%%%%%%%%%%%%%%%%%%%%%%%%%%%%%%%%%%%%%%%%%%%%%%%%%%%%%%%%%
\begin{figure}[t!]
\centerline{
\resizebox{8cm}{!}{ \includegraphics*{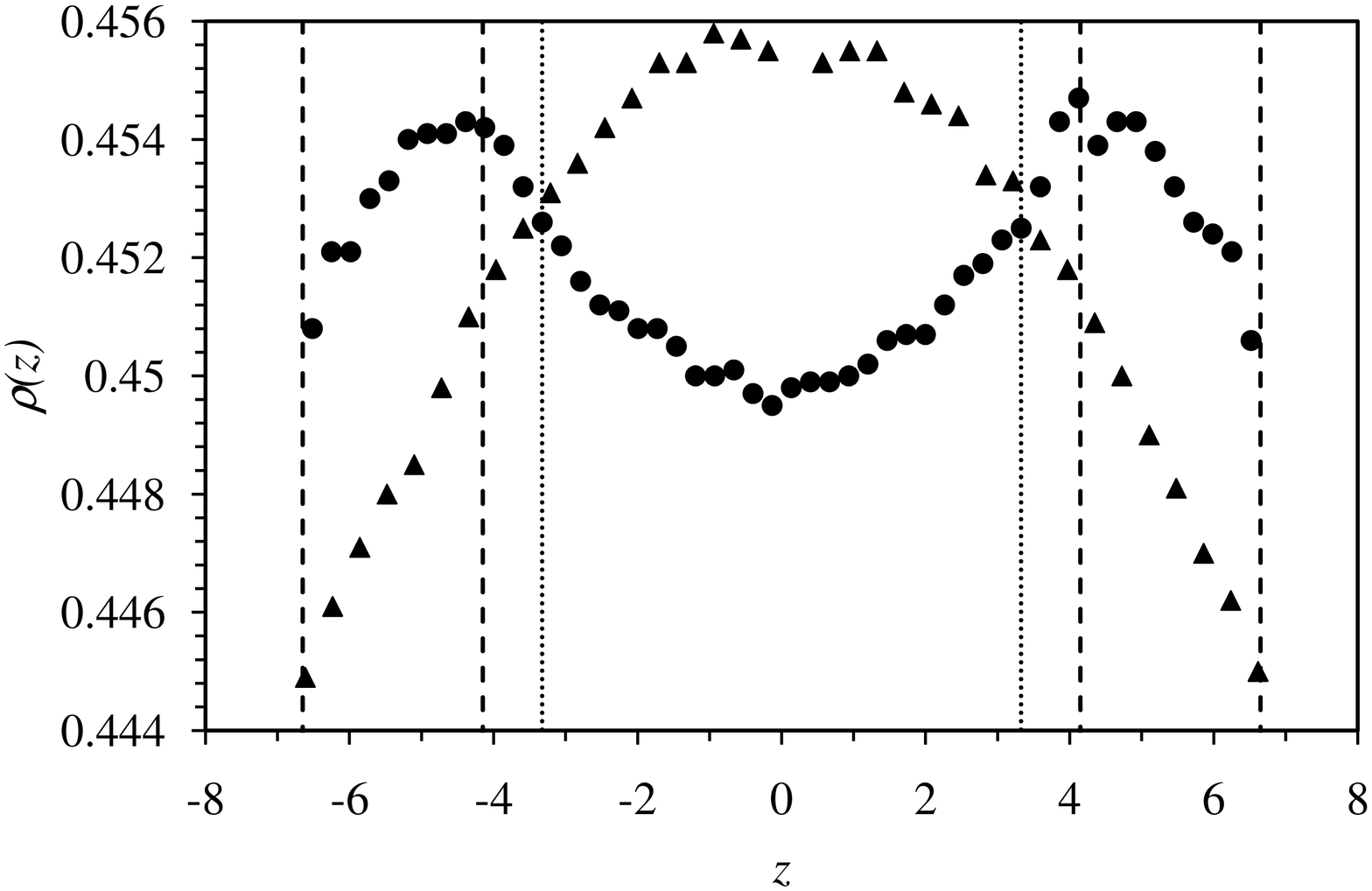} } }
\caption{\label{Fig:rho(z)}
Density profile,
with the state point, symbols, and lines
as in the preceding figure.
}
\end{figure}
%%%%%%%%%%%%%%%%%%%%%%%%%%%%%%%%%%%%%%%%%%%%%%%%%%%%%%%%%%%%%%%%%%

Figure~\ref{Fig:rho(z)} shows the density profile
for the uniform and slab parameter cases.
In the uniform $n=0$ case,
the density depletion with increasing distance from the center
appears to correlate with the increase
in the mean speed $|\overline v_x(z)|$
and hence the kinetic energy $\overline{\cal K}(z)$.
The  explanation is that it is entropically favorable
to transfer  peripheral particles to the center
where they have lower speed
because this releases energy to the reservoir
and hence the entropy associated with the phase space point is increased.
%Notice also how the periodic boundary conditions
%induce a cusp and sharp density depletion right at the boundary.
For the uniform parameter case,
it was found that the peak of the density profile
increased with increasing $\theta$ at constant $\gamma$.
This is consistent with the slab result,
where the  density is enhanced in the slab region
compared to the adiabatic region.
Undoubtedly these density inhomogeneities contribute
to the sensitivity of the shear viscosity to the
value of the fluctuation-dissipation parameter.
Given the 1--2\% difference in temperature and density profiles
between the uniform and slab application
of the fluctuation-dissipation parameter,
it is unrealistic to expect the two methods
to produce viscosities in better agreement than this.
Ultimately the two formulations model different physical situations,
with the uniform case applicable to the small size asymptote,
and the boundary slab applicable to the size-dependent adiabatic case.

%%%%%%%%%%%%%%%%%%%%%%%%%%%%%%%%%%%%%%%%%%%%%%%%%%%%%%%%%%%%%%%%%%
\begin{figure}[t!]
\centerline{
\resizebox{8cm}{!}{ \includegraphics*{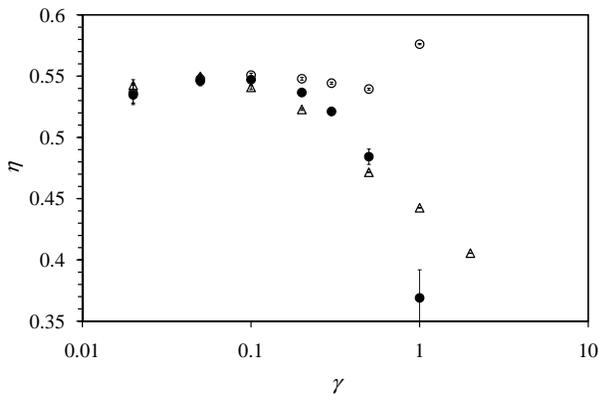} } }
\caption{\label{Fig:eta-gamma-.452}
Shear viscosity versus shear rate at $T=2$, $\rho=0.452$,
using the constrained equations of motion.
The triangles use $n=0$, $\theta=10$
(except $\gamma=2$, which uses $\theta=50$).
The circles use the boundary slab with $\theta=10$
(except  $\gamma=0.5$, which uses $\theta=20$,
and $\gamma=1$, which uses $\theta=100$).
The empty symbols are the measured viscosity,
and the filled  symbols have been temperature corrected (see text)
using the measured
$\partial \eta/\partial T = 0.0800 \pm 0.0089$.
}
\end{figure}
%%%%%%%%%%%%%%%%%%%%%%%%%%%%%%%%%%%%%%%%%%%%%%%%%%%%%%%%%%%%%%%%%%

Figure~\ref{Fig:eta-gamma-.452}
shows the viscosity as a function of shear rate
at the same state point as above,
$T=2$, $\rho=0.452$.
The constrained equations of motion were used
with uniform ($n=0$, triangles)
and boundary slab (circles) application
of the fluctuation-dissipation forces.
Except at the highest shear rates,
the same value of the fluctuation-dissipation parameter was used.
At low shear rates the uniform and the slab results
asymptote to the same constant viscosity, $\eta = 0.54 \pm 0.01$.
It can be seen that for the uniform case,
there is a noticeable shear thinning
for shear rates $\gamma \agt 0.1$.
Recall that for this state point
the  limit for negligible temperature change, Eq.~(\ref{Eq:low-gamma-lim}),
is $\gamma_\mathrm{low} = 0.61$.

For the boundary slab case,
the shear viscosity was obtained by dividing
the average $xz$ component of the diffusive momentum tensor
in the central half of the sub-system
by the gradient in the velocity,
$\partial \overline v_x(z)/\partial z$,
actually in that region.
At this and the higher densities below,
the latter was almost always higher than the applied shear rate,
increasingly so as the applied shear rate was increased.

For the boundary slab case,
the kinetic temperature in  the adiabatic central region
increased significantly at high shear rates,
and a correction was used to facilitate comparison
with the uniform fluctuation-dissipation parameter case.
The temperature corrected viscosity is
$\eta_\mathrm{corr} = \eta - \eta' [ \overline T_\mathrm{s} - T]$,
where the temperature derivative $\eta' \equiv \partial \eta/\partial T$
is obtained by finite difference of two simulations at $T=2 \pm 0.1$.
In this case $\eta' = 0.0800 \pm 0.0089$
was obtained at $\gamma=0.1$ with $\theta=10$ and $n=0$.
Also, $\eta$ is the average viscosity
and
$\overline T_\mathrm{s}$ is the average kinetic temperature
in the central adiabatic half.

This temperature correction is problematic
at high shear rates for three reasons.
First, at high shear rates
the kinetic temperature is not the actual temperature,
Eq.~(\ref{Eq:EquiThm-Shear}).
Second, no correction is made
for the density difference.
And third, the linear correction is unreliable
for large temperature shifts.

Despite these concerns,
it is clear that the temperature difference
accounts for much of the difference between the viscosity measured
in the uniform and in the slab cases
at low to moderate shear rates, $\gamma \alt 0.5$.
At the highest shear rate $\gamma=1$,
the actual viscosity in the slab case is larger than that
in the low shear rate limit,
whereas the temperature-corrected viscosity is smaller than the limiting value.
One could describe the raw data as shear thickening,
and the corrected data as shear thinning.
In this case it is clear that the  shear thickening
comes entirely from the temperature increase due to viscous heating
in the adiabatic region, since at this state point
$ \partial \eta/\partial T > 0$.

%It is worth mentioning that
%the $n=0$ $\theta=10$ estimate $\eta=0.46715 \pm 0.00069$ at $\gamma=0.5$
%is comparable to the  estimate of Hess\cite{Hess02}
%(NEMD SLLOD, $\gamma=0.5$, thermostat coupling of 20)
% $\eta = 0.462 \pm 0.008$.
%It is possible that the thermostat has negligible influence
%at this shear rate;
%it is also possible that the agreement may be
%no more than a fortuitous coincidence.

%%%%%%%%%%%%%%%%%%%%%%%%%%%%%%%%%%%%%%%%%%%%%%%%%%%%%%%%%%%%%%%%%%
\begin{figure}[t!]
\centerline{
\resizebox{8cm}{!}{ \includegraphics*{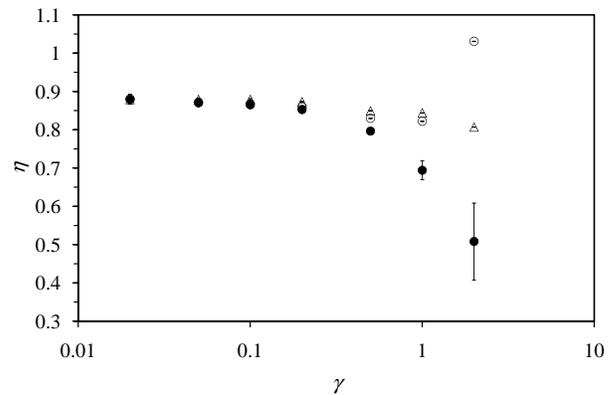} } }
% Projects/Shear18/d4shear.xlsx:chart5
\caption{\label{Fig:eta-gamma-.6}
Shear viscosity versus shear rate at $T=2$, $\rho=0.6$,
using the constrained equations of motion.
The triangles use $n=0$, $\theta=5$,
and the circles use the boundary slab with $\theta=10$.
The empty symbols are the measured viscosity,
and the filled  symbols have been temperature corrected
using the measured
$\partial \eta/\partial T = 0.057 \pm  0.011 $
(obtained at $\gamma=0.1$ with $\theta=5$ and $n=0$).
}
\end{figure}
%%%%%%%%%%%%%%%%%%%%%%%%%%%%%%%%%%%%%%%%%%%%%%%%%%%%%%%%%%%%%%%%%%

%%%%%%%%%%%%%%%%%%%%%%%%%%%%%%%%%%%%%%%%%%%%%%%%%%%%%%%%%%%%%%%%%%
\begin{figure}[t!]
\centerline{
\resizebox{8cm}{!}{ \includegraphics*{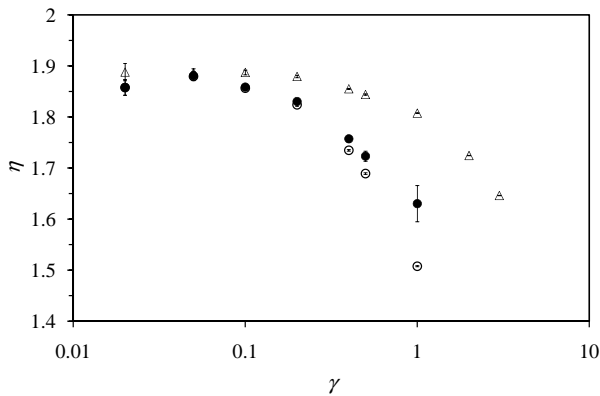} } }
\caption{\label{Fig:eta-gamma-.8}
Shear viscosity versus shear rate at $T=2$, $\rho=0.8$,
using the constrained equations of motion.
The triangles use $n=0$, $\theta=2$
(except $\gamma= $ 1 and 2, which use $\theta=10$,
and $\gamma=3$, which uses  $\theta=50$).
The circles use the boundary slab with $\theta=20$
(except $\gamma=$ .05 and .02, which use $\theta=2$).
The empty symbols are the measured viscosity,
and the filled  symbols have been temperature corrected
using the measured
$\partial \eta/\partial T = -0.064 \pm 0.019$
(obtained at $\gamma=0.1$ with $\theta=2$ and $n=0$).
}
\end{figure}
%%%%%%%%%%%%%%%%%%%%%%%%%%%%%%%%%%%%%%%%%%%%%%%%%%%%%%%%%%%%%%%%%%

Figures~\ref{Fig:eta-gamma-.6} and \ref{Fig:eta-gamma-.8}
shows the viscosity as a function of shear rate
at $T=2$ and at $\rho=0.6$ and $\rho=0.8$, respectively.
The limiting shear rate for negligible temperature increase,
Eq.~(\ref{Eq:low-gamma-lim}),
is $\gamma_\mathrm{low} = 0.73$ for $\rho=0.6$
(using  $\lambda=4.1$\cite{AttardV} and $\eta=0.88$),
and $\gamma_\mathrm{low} = 0.70$ for $\rho=0.8$
(using  $\lambda=7.2$\cite{AttardV} and $\eta=1.9$).

Again both the uniform and the slab fluctuation-dissipation methods
agree on the low shear rate asymptotes.
The low shear rate viscosity evidently increases with increasing density.
Further both methods,
at least initially, predict shear thinning for both densities,
although the quantitative agreement between the two models is limited
at higher shear rates.

At $\rho=0.6$, the raw boundary slab results show shear thickening
at higher shear rates due to the increasing temperature
in the adiabatic central region and the positive value
of the temperature derivative of the viscosity.
Correcting for this temperature effect gives shear thinning behavior.

Interestingly enough, at the higher density $\rho=0.8$ the temperature
derivative of the viscosity is negative,
and both the raw and the corrected results show shear thinning.
The temperature corrected boundary slab results again lie closer to
the uniform fluctuation-dissipation parameter results,
which confirms that much of the difference between the two
can be attributed to the higher temperature in the adiabatic region.

For most of the results at  $\rho=0.8$,
increasing the value of $\theta$ in the uniform constrained case
increased the value of $\eta$ at a given $\gamma$.
This is the opposite behavior to that observed
in the above results at the lower density.
Possibly the sign of this effect depends
on the sign of the temperature derivative of the viscosity,
since increasing $\theta$ generally decreases the kinetic temperature.

All of the results presented above and below are for a time step of
$\Delta_t = 0.01 $.
It should be mentioned that halving this to
$\Delta_t = 0.005$ at $\rho=0.6$ and $\gamma=2$
changes  the various averages by 0.1--0.5\%,
which is about an order of magnitude larger
than the typical statistical error.
This systematic error is less at lower shear rates.

The quantitative effects of finite size were estimated
at $T=2$, $\rho=0.6$, and $\gamma=1$,
in the uniform $n=0$ case (1000 particles).
At $\theta=5$,
changing $L_z$ from 11.9 to 15.4
decreases $T^\mathrm{kin}$ by 0.2\% and decreases $\eta$ by 1\%.
%For $n=0$ and $\theta=10$ the same relative changes hold.
Changing  $\theta$ from 5 to 10 at $L_z=11.9 $
changes $T^\mathrm{kin}$ from 2.39 to 2.20, and decreases $\eta$ by 5\%.

%%%%%%%%%%%%%%%%%%%%%%%%%%%%%%%%%%%%%%%%%%%%%%%%%%%%%%%%%%%%%%%%%%
\begin{figure}[t!]
\centerline{
\resizebox{8cm}{!}{ \includegraphics*{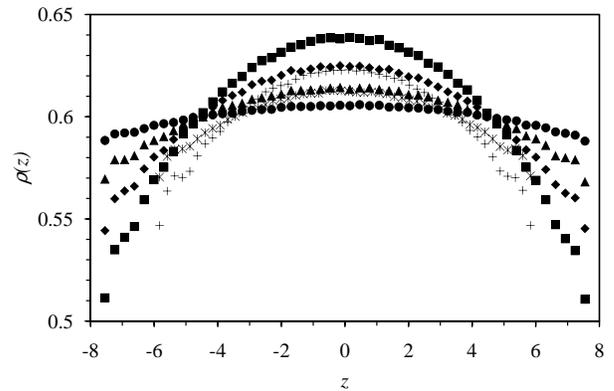} } }
% Projects/Shear18/d4shear.xlsx:chart7
\caption{\label{Fig:rho(z)-.6}
Density profile at $T=2$, $\rho=0.6$,
and $\gamma=1$ using  the constrained equations of motion
with a uniform parameter $n=0$.
The geometrical shapes are for $L_z=15.4$
and, from bottom to top at the center,
$\theta = $ 2, 5, 10, and 20.
The characters are for  $L_z=11.9$
and, from bottom to top at the center,
$\theta = $ 5 and 10.
}
\end{figure}
%%%%%%%%%%%%%%%%%%%%%%%%%%%%%%%%%%%%%%%%%%%%%%%%%%%%%%%%%%%%%%%%%%

Density profiles for the uniform $n=0$ case
are shown in Fig.~\ref{Fig:rho(z)-.6}.
The density profile is of the form
$\rho(z) = \rho(0) - 12\{ \rho(0) - \overline \rho \} z^2/L_z^2$,
with the central density $\rho(0)$ being independent of system size
and increasing with increasing fluctuation-dissipation parameter.
The corresponding temperature profiles are uniform,
show a similar independence from system size,
and $T(z)-T$ decreases
with increasing fluctuation-dissipation parameter value.
%The inhomogeneous density profile appear to be induced
%in part by the periodic sliding brick boundary conditions.
%From symmetry, any inhomogeneity has to be even in $z$,
%$\rho(z) = \rho(-z)$.
%Hence there must be a discontinuity in the derivative
%at the periodic boundary,
%$\rho'(L_z/2 + \varepsilon) = -\rho'(L_z/2-\varepsilon) $,
%$\varepsilon\rightarrow 0$.
%This cusp is presumably unfavorable and leads to a density depletion
%at the periodic boundary,
%and a corresponding maximum density at $z=0$.
More broadly,
the magnitude of the density inhomogeneity
increases with shear rate and with fluctuation-dissipation parameter.

%%%%%%%%%%%%%%%%%%%%%%%%%%%%%%%%%%%%%%%%%%%%%%%%%%%%%%%%%%%%%%%%%%
\begin{figure}[t!]
\centerline{
\resizebox{8cm}{!}{ \includegraphics*{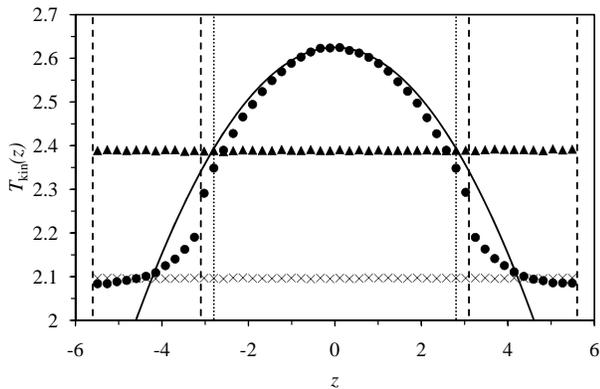} } }
\caption{\label{Fig:T(z)-.8}
Kinetic temperature profile at $T=2$, $\rho=0.8$,
and $\gamma=0.5$ using  the constrained equations of motion.
The circles are the boundary slab case with $\theta = 20$,
and the triangles and crosses are the uniform parameter case $n=0$
with $\theta=2$ and $\theta=10$, respectively.
The solid curve is the expected adiabatic temperature profile,
Eq.~(\ref{Eq:T(z)}) using $\eta=1.7$ and $\lambda=7.2$,\cite{AttardV}
with $T(0)$ fitted.
The dashed lines delineate the boundary slab,
and the dotted lines delineate the central half of the sub-system.
The system is periodic beyond the plotted data.
}
\end{figure}
%%%%%%%%%%%%%%%%%%%%%%%%%%%%%%%%%%%%%%%%%%%%%%%%%%%%%%%%%%%%%%%%%%

The kinetic temperature profiles shown in Fig.~\ref{Fig:T(z)-.8}
confirm the uniformity for the $n=0$ case,
and the increasing efficacy of the fluctuation-dissipation parameter
with increasing magnitude.
For the boundary slab case,
one again sees that the continuum expression Eq.~(\ref{Eq:T(z)})
is applicable in the central adiabatic region.
The small discrepancy between this expression
and the measured values
can either be taken to indicate small local changes in $\eta$ or $\lambda$
as a result of the inhomogeneity in $T(z)$ or $\rho(z)$,
or else it indicates that
the difference between the kinetic temperature and the actual temperature,
the neglected higher order contributions in Eq.~(\ref{Eq:T(z)}),
are important at this high shear rate.

%%%%%%%%%%%%%%%%%%%%%%%%%%%%%%%%%%%%%%%%%%%%%%%%%%%%%%%%%%%%%%%%%%
\begin{figure}[t!]
\centerline{
\resizebox{8cm}{!}{ \includegraphics*{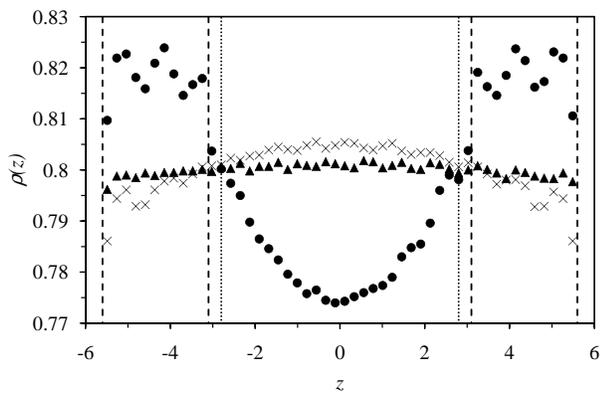} } }
\caption{\label{Fig:rho(z)-.8}
Density profile,
with the state point, symbols, and lines
as in the preceding figure.
}
\end{figure}
%%%%%%%%%%%%%%%%%%%%%%%%%%%%%%%%%%%%%%%%%%%%%%%%%%%%%%%%%%%%%%%%%%

The density profiles shown in Fig.~\ref{Fig:rho(z)-.8}
are qualitatively similar to those shown for the lower density case,
Fig.~\ref{Fig:rho(z)}.
The sharp density depletion at the periodic boundary
in the uniform fluctuation-dissipation parameter case, $n=0$,
induces an adjacent clear local maximum or density oscillation
at this high density.
The behavior of $\rho(z)$ for the boundary slab case
is qualitatively different, with a minimum at $z=0$.
It can be argued that the density enhancement in the slab region
is consistent with the uniform parameter case
in that a large fluctuation-dissipation parameter
apparently favors a higher density.

At $T=2$, $\rho=0.8$, $\gamma=0.4$
for the boundary slab case with  $\theta = 20$,
in the central adiabatic region
the simulations show that
$T_{x}^\mathrm{kin}=2.36450 \pm 0.00065 $,
$T_{y}^\mathrm{kin}= 2.35002 \pm 0.00067  $,
and $T_{z}^\mathrm{kin}= 2.33883 \pm 0.00055$.
(The change upon halving the time step to $\Delta_t=.005$ in this case
is smaller than the statistical error.)
Evidently then
$ T_{x}^\mathrm{kin} > T_{y}^\mathrm{kin} > T_{z}^\mathrm{kin}$.
This ordering agrees with Hoover \emph{et al.},\cite{Hoover08}
and it appears significant compared to any possible systematic error.
%(At $\rho=0.6$ and $\gamma=2$
%the systematic error due to the finite time step
%was estimated at 0.1--0.5\%.)
Also, for the pressure tensor in the same region,
$ P_{xx} = 7.0489 \pm 0.0019 $,
$ P_{yy} = 6.9912 \pm 0.0019$, and
$ P_{zz} = 6.9897 \pm 0.0017 $.
Hence
$  P_{xx} > P_{yy} = P_{zz}$.
This ordering agrees with Hoover \emph{et al.}\cite{Hoover08}
Halving the time step to $\Delta_t=.005$  changes
these to
$ P_{xx} = 6.9923 \pm 0.0020 $,
$ P_{yy} = 6.9362 \pm 0.0020$, and
$ P_{zz} = 6.9322 \pm 0.0017 $.

At the same state and shear point for the uniform case $n=0$, $\theta=2$,
the simulations show that for the whole system
$T_{x}^\mathrm{kin}= 2.25865 \pm 0.00033$,
$T_{y}^\mathrm{kin}=  2.24732 \pm 0.00041  $, and
$T_{z}^\mathrm{kin}= 2.25232 \pm 0.00037$.
Evidently then
$ T_{x}^\mathrm{kin} \agt  T_{z}^\mathrm{kin} > T_{y}^\mathrm{kin} $.
(The change upon halving the time step to $\Delta_t=.005$ in this case
is smaller than the statistical error.)
In the central half of the sub-system,
$ P_{xx} = 7.0768 \pm 0.0018 $,
$ P_{yy} =  7.0237 \pm 0.0017$, and
$ P_{zz} =  7.0529 \pm 0.0016$.
Hence
$ P_{xx} > P_{zz} > P_{yy}$.
(In the whole system,
$ P_{xx} =  7.0304 \pm 0.0014$,
$ P_{yy} =  7.0038 \pm 0.0014$, and
$ P_{zz} =  7.0320 \pm 0.0014$.)
Halving the time step to $\Delta_t=.005$  changes
these to
$ P_{xx} =  7.0267 \pm 0.0019$,
$ P_{yy} =  6.9758 \pm 0.0019$, and
$ P_{zz} =  7.0054 \pm 0.0018$,
for the central half,
and to
$ P_{xx} =  6.9805 \pm 0.0017$,
$ P_{yy} =  6.9566 \pm 0.0018$, and
$ P_{zz} =  6.9850 \pm 0.0017$,
for the whole system.)
It should be noted that for the $n=0$ case,
the shear rate is uniform throughout the system,
but the density is higher and varies more slowly
in the center of the system
compared to the periphery.

%The thermal conductivity of a Lennard-Jones fluid
%at $T=2$ and $\rho=0.8$ is about $\lambda=7.2$.\cite{AttardV}
%At $\rho=0.5$ it is 2.5 and at $\rho=0.3$ it is 1.5.
%Hence at $\rho=0.6$ it can be interpolated to be $\lambda=4.1$,
%from the results in Ref.~[\onlinecite{AttardV}].

%%%%%%%%%%%%%%%%%%%%%%%%%%%%%%%%%%%%%%%%%%%%%%%%%%%%%%%%%%%%%%%%%%%%%%%%%%
%
\section{Discussion and Conclusions}
\setcounter{equation}{0} \setcounter{subsubsection}{0}
%
%%%%%%%%%%%%%%%%%%%%%%%%%%%%%%%%%%%%%%%%%%%%%%%%%%%%%%%%%%%%%%%%%%%%%%%%%%

The main conclusions to be drawn from this work may be divided
into those specific for shear flow in simple fluids,
and those applicable to computer simulations
and non-equilibrium systems more generally.

The specific results obtained here
show that Lennard-Jones fluids are shear thinning,
at least when the kinetic temperature is held close to that of the reservoir.
Conversely,
when the kinetic temperature of the sub-system is allowed to rise
with increasing shear rate,
as it does in steady state, boundary driven flow,
then the fluid can be shear thickening
if the temperature derivative of the viscosity is positive,
as it is at low to moderate densities.

One of the results with more general implications
is the demonstrated sensitivity of the shear viscosity to the value
of the fluctuation-dissipation parameter,
which complements earlier studies
that have shown its dependence
on the specific non-equilibrium equations of motion,
the deterministic thermostat,
or the boundary conditions.
\cite{Hoover08,Khare97,Delhommelle03,Berro11}
%,Loose89,Loose92,Bagchi96,Evans86,Padilla95,Travis96}
The present stochastic, dissipative equations of motion
provide more certainty in the results than these earlier studies
because they are  consistent
with the non-equilibrium probability distribution,
and because they do not require the sub-system temperature
(other methods  assume both
that the  sub-system temperature is given
by an equilibrium phase space function,
and that it is uniform).

It was shown here that imposing a constraint on the equations of motion
reduces the sensitivity %of the simulated viscosity
to the value of the fluctuation-dissipation parameter,
which thereby increases the confidence
in the quantitative value of the simulated viscosity.
The general formulation of the constraint given here
allows this technique to be applied to general non-equilibrium systems.

With the constraint
the present results can be considered quantitative
for low to moderate shear rates, $\gamma \alt 0.2$,
but only qualitatively reliable for high shear rates.
Fortunately the shear thinning behavior begins
in the moderate shear rate regime,
and so that general conclusion can be considered reliable.
One limitation of the stochastic, dissipative equations of motion
is that they are exact only to linear order
in the fluctuation-dissipation parameter,
which accounts for the dependence of the viscosity on
the large values of  $\theta$ required at high shear rates to dissipate
the viscous heat.

A second problem at high shear rates is with the meaning of temperature.
Equating the kinetic temperature (or the Rugh temperature)
to the sub-system temperature
is not valid beyond linear order in the shear rate,
Eq.~(\ref{Eq:EquiThm-Shear}).
This creates uncertainty in those algorithms that use deterministic
thermostats to control the sub-system temperature.
Although the reservoir temperature,
which is used in the present stochastic, dissipative equations of motion,
is well-defined,
in practice one often judges the efficacy of the value
of the  fluctuation-dissipation parameter at a given shear rate
by the consonance between the sub-system kinetic temperature
and the reservoir temperature.
Although this is arguably irrational,
it is difficult to suggest an alternative way
of choosing an appropriate value of $\theta$.
One must concede that
there is a degree of uncertainty
in the results of the present SDMD method at high shear rates.

A third limitation is that high shear rates
magnify the difference between the uniform application
of the fluctuation-dissipation parameter,
and the boundary driven flow where the parameter
is applied  only in the boundary region.
In the latter model,
the motion is adiabatic in the central region,
which consequently attains a higher temperature than the reservoir
as heat can only dissipate via the boundaries.
As Hoover \emph{et al.}\cite{Hoover08} point out,
this effect increases with system size.
This,
together with the similarly size-dependent
density inhomogeneity induced by the shear flow
(Figs~\ref{Fig:rho(z)}, \ref{Fig:rho(z)-.6}, \ref{Fig:rho(z)-.8}),
makes the shear viscosity dependent on the system size in the direction
of the shear gradient.

The power of Gibbs' formulation of equilibrium thermodynamics
in terms of reservoirs
is that it allows the sub-system to be described in molecular detail,
as functions of the intensive thermodynamic properties of the reservoir.
The useful properties of the sub-system can almost always be expressed
as intensive variables that are independent of the sub-system size.
(Of course equilibrium systems that are inhomogeneous,
due to an external potential or to their  proximity to a surface,
are an exception.)
Moreover, the conjugate intensive variables to the material exchangeable
with the reservoir have the same values as those of the reservoir.
These are extremely useful results
because it means the simulated or measured properties
of the sub-system are a function solely of the specified
thermodynamic state point,
and they can be used in a local sense in any system
composed of the same material and in the same state
irrespective of the size of the system.

The generic nature of reservoirs is reflected
in the abstract formulation of the stochastic dissipative equations of motion.
Irrespective of the chemical composition,
physical state, or geometry of the reservoir,
the formalism guarantees the correct equilibrium probability distribution
corresponding to the exchangeable variables.
Similarly,
the stochastic dissipative equations of motion
yield the correct probability for the sub-system
insensitive to the precise value
of the fluctuation-dissipation parameter or to its place of application.
The abstract and generic nature of the formalism
can be expected to carry over to non-equilibrium systems
in the linear regime.

The size-dependence of the shear viscosity identified here
diminishes to some extent the utility of the reservoir formalism
in the case of non-linear shear flow.
In the limit of low shear rate or small sub-system size
the difference between boundary-driven flow
and a uniformly applied parameter vanishes,
the shear viscosity becomes a normal intensive variable,
and the usual rules apply.
However at high shear rates one is faced with difficult questions
when it actually comes to using
the shear viscosity in the hydrodynamic equations.
Arguably,
one should use a local value of the viscosity
that depends on the local shear rate and temperature.
Such a local value corresponds to the small-size limit,
and in this case the viscosity obtained
at this state point with the uniform parameter applies.
%If the shear thinning behavior identified here were applied locally,
%it could possibly account for the onset of turbulence,
%a quantitative version of the yield stress model proposed by Paglietti.
%\cite{Paglietti17}

There remain a number of unanswered questions regarding shear flow,
and simulations of non-equilibrium systems more generally.
The present stochastic, dissipative equations of motion
are extremely convenient,
but the sensitivity to the fluctuation-dissipation parameter
at high shear rates,
even with the constraint imposed,
reflects the physical sensitivity to the nature of the reservoir
in the non-linear non-equilibrium regime,
and this creates doubt as to the meaning of the results.
One way forward is possibly to expand the transition probability
beyond first order in the parameter.
Alternatively,  one could perform Monte Carlo simulations
based on the non-equilibrium probability distribution,
as has previously been done for heat flow\cite{AttardV}
and for driven Brownian motion.\cite{Attard09}
The non-equilibrium Monte Carlo (NEMC) method
is somewhat inefficient computationally,
but it has the advantage that versions can be formulated
without invoking the fluctuation-dissipation parameter.
However NEMC implicitly relies upon the reservoir formalism,
so it is not at all clear that it would resolve all the ambiguities
discussed here in the non-linear non-equilibrium regime.

%\newpage %$\;$ \newpage
%\section{References}
%%%%%%%%%%%%%%%%%%%%%%%%%%%%%%%%%%%%%%%%%%%%%%%%%%%%%%%%%%%%%%%%%%%%%%%%%%

\end{document}